\documentclass{emulateapj}
\usepackage{natbib}
\begin{document}

\title{Protoplanetary Disk Properties in the Orion Nebula Cluster:
  Initial Results from Deep, High-Resolution ALMA Observations}

\author{J. A.  Eisner\altaffilmark{1}, H. G. Arce\altaffilmark{2},
  N. P. Ballering\altaffilmark{1},  J. Bally\altaffilmark{3},
 S. M. Andrews\altaffilmark{4},  R. D. Boyden\altaffilmark{1}, 
  J. Di Francesco\altaffilmark{5}, 
  M. Fang\altaffilmark{1}, D. Johnstone\altaffilmark{5},
  J. S. Kim\altaffilmark{1}, R. K. Mann\altaffilmark{5}, B. Matthews\altaffilmark{5},
  I. Pascucci\altaffilmark{6}, 
  L. Ricci\altaffilmark{7}, P. D. Sheehan\altaffilmark{8}, J. P. Williams\altaffilmark{9}}
\altaffiltext{1}{Steward Observatory, University of Arizona, 933 North
  Cherry Avenue, Tucson, AZ 85721, USA}
\altaffiltext{2}{ Department of Astronomy, Yale University, New Haven,
CT 06520}
\altaffiltext{3}{ Department of Astrophysical and Planetary Sciences,
 University of Colorado, UCB 389, Boulder, CO 80309, USA}
\altaffiltext{4}{Harvard-Smithsonian Center for Astrophysics, 60
  Garden Street, Cambridge, MA 02138, USA}
\altaffiltext{5}{NRC Herzberg Astronomy and Astrophysics, 5071 West
  Saanich Road, Victoria, BC, V9E 2E7, Canada}
\altaffiltext{6}{Lunar and Planetary Laboratory, University of
  Arizona, Tucson, AZ 85721}
\altaffiltext{7}{Department of Physics and Astronomy, California State
  University Northridge, 18111 Nordhoff St, Northridge, CA 91330, USA}
\altaffiltext{8}{Homer L. Dodge Department of Physics and Astronomy,
  University of Oklahoma, 440 W. Brooks Street, Norman, Oklahoma
  73019}
\altaffiltext{9}{Institute for Astronomy, University of Hawaii, 2680
  Woodlawn Drive, Honolulu, HI 96822}
\email{jeisner@email.arizona.edu}


%
\slugcomment{Accepted for Publication in ApJ}

\begin{abstract}
We present ALMA 850 $\mu$m  continuum observations of the Orion
Nebula Cluster that provide
the highest angular resolution ($\sim 0\rlap{.}''1 \approx 40$ AU) and
deepest sensitivity ($\sim 0.1$ mJy) of the region
to date.  We mosaicked a field containing  $\sim 225$
optical or near-IR-identified young stars,  $\sim 60$ of which are also
optically-identified ``proplyds''.  We detect
continuum emission at 850 $\mu$m towards $\sim 80\%$ of the
proplyd sample, and $\sim 50$\% of the larger sample of
previously-identified cluster members.
Detected objects have fluxes of $\sim 0.5$--80 mJy.  We
remove sub-mm flux due to free-free emission in some objects,
leaving a sample of sources detected in dust emission.
Under standard assumptions of isothermal, optically thin
disks, sub-mm fluxes correspond to dust masses of $\sim 0.5$ to 80
Earth masses.  We measure
the distribution of disk sizes, and find that disks in this region are
particularly compact.  Such compact disks are likely to be
significantly optically thick.  The distributions of sub-mm flux and
inferred disk size indicate smaller, lower-flux disks than in 
lower-density star-forming regions of similar age.  
Measured disk flux is
correlated weakly with stellar mass, contrary to studies in
other star forming regions that found steeper correlations.  
We find a
correlation between disk flux and distance from the massive star
$\theta^1$ Ori C, suggesting that disk properties in this region are
influenced strongly by the rich cluster environment.
\end{abstract}

\section{Introduction}
Protoplanetary disks are the birth-sites of planetary systems, and the
properties of disks relate directly to the planets that may
potentially form.  In particular, disk mass constrains the mass budget
for planet formation, and hence the mass distribution of planets.
The minimum-mass solar nebula (MMSN) needed to form the planets in our
Solar System was likely between 0.01 M$_{\odot}$ and 0.1 M$_{\odot}$
\citep[e.g.,][]{WEID+77,DESCH07}.  

To build giant
planets on timescales shorter than inferred gas disk lifetimes
\citep[$\sim$2--5 Myr; e.g.,][]{FEDELE+10,INGLEBY+12,DARIO+14}, 
models require initial disk masses (including both gas and solids) 
$\ga 0.01$ M$_{\odot}$
\citep[e.g.,][]{HAYASHI81,ALIBERT+05}.  Meteoritic evidence in our
solar system suggests that the core of Jupiter likely formed on a
much shorter timescale, $<1$ Myr \citep{KRUIJER+17,DESCH+17}.
Protoplanetary disks with cleared gaps or holes around stars aged $\la
1$ Myr also suggest rapid formation of giant planets
\citep[e.g.,][]{ALMA+15,DONG+15,SE17}.  
Higher disk masses accelerate planet formation timescales (since
collision rate scales with surface density).  The apparently rapid
growth of $>10$ Earth-mass cores suggests that required initial
disk masses may be on the high end of the MMSN range.

To constrain the mass of solid particles in disks, one must observe at
long wavelengths.  At short wavelengths ($\lambda \la 10$ $\mu$m), the
dust in protoplanetary disks is optically thick even for solid masses
$<10^{-8}$ M$_{\odot}$.   Furthermore, longer-wavelength observations
are sensitive to larger particles, and thus mass estimates are
somewhat less affected by potential particle growth.

Historically, disk masses have been constrained by converting observed
mm-wave or sub-mm-wave emission into mass under a range of assumptions
\citep[e.g.,][]{BECKWITH+90}.  Since continuum opacity in disks is
dominated by dust, investigators generally began by estimating dust masses.
A key assumption is that the dust is
optically thin, and so observed emission is directly correlated with
the mass of solids.  One must also choose dust opacities and temperatures.
To compute the total, gas+dust mass, an additional assumption about
the dust-to-gas ratio is required.  

All of these assumptions are potentially problematic.  Compact, but
massive protoplanetary disks are likely to be substantially optically
thick even at sub-mm wavelengths \citep[see, e.g.,][]{WU+17}.
Opacities and
temperatures can vary within individual systems, and from star to star
\citep[e.g.,][]{ANDREWS+13}.
While gas-to-dust ratios are often assumed to be the ISM value of 
100 for all systems, recent observations and modeling suggest this may
often be incorrect \citep[e.g.,][]{WB14,EISNER+16,MIOTELLO+17}.

To obtain more accurate disk masses, spatially resolved imaging is
needed.  If disks are resolved, we can constrain optical depth, as
well as radial density and temperature profiles.  Moreover, with
sufficient resolution we can measure the mass within the inner 
$\sim 30$ AU planet-forming region, providing direct constraints on
planet formation theory.

Disk masses have been well-studied previously in low-density star
forming regions
\citep[e.g.,][]{AW05,AW07,ANDREWS+13,WILLIAMS+13,CARPENTER+14,ANSDELL+15,ANSDELL+16,ANSDELL+17,PASCUCCI+16,BARENFELD+16,BARENFELD+17,LAW+17}.
Spatially resolved observations of disks in these regions have
constrained relationships between protoplanetary disk and stellar
properties.   For example, studies of disks in the Taurus,
Ophiuchus, Lupus,
Chameleon 1, and Upper Sco star-forming regions have shown a
dependence of disk flux on stellar mass
\citep{ANDREWS+13,ANSDELL+16,PASCUCCI+16,BARENFELD+16}, a relation
that appears to steepen with age \citep{PASCUCCI+16}.
Sub-mm flux also appears to correlate strongly with disk size in these
regions \citep{TRIPATHI+17,TAZZARI+17}. 

Most stars in the Galaxy do not form in low-density star-forming
regions.  Rather, most stars form in rich clusters like the Orion Nebula
\citep{LADA+91,LSM93,CARPENTER00,LL03}.  Isotopic abundances and
dynamical signatures in 
our Solar System suggest that it, too, may have formed in 
a dense, Orion-like environment \citep[e.g.,][]{HD05,WG07,ADAMS10,DK12}. 
Outflow feedback, UV radiation, stellar winds, and tidal encounters in
rich clusters produce a profoundly different environment than that
found in lower-density regions
\citep[e.g.,][]{SC01,TB05b,ADAMS+06,HOLDEN+11}.
Expanding sub-millimeter continuum surveys to include rich clusters allows the 
determination of protoplanetary disk properties and evolutionary timescales in
typical star (and planet) formation environments.  

Rich clusters are relatively challenging to observe because of their
distances and high stellar density. High angular resolution
and sensitivity are required.  Only a handful of nearby rich clusters have
been observed to date: the Orion Nebula cluster
\citep{MLL95,BALLY+98,WAW05,EC06,EISNER+08,EISNER+16,MW09,MW10,MANN+14}, 
IC 348 \citep{CARPENTER02,CIEZA+15}, and NGC 2024 \citep{EC03,MANN+15}. 
These existing surveys have detected very few disks with fluxes high enough
to suggest $\ga 0.01$--0.1 M$_{\odot}$ of material \citep[e.g.,][]{EISNER+08,EISNER+16}.


\epsscale{1.0}
\begin{figure*}[tbhp]
\plotone{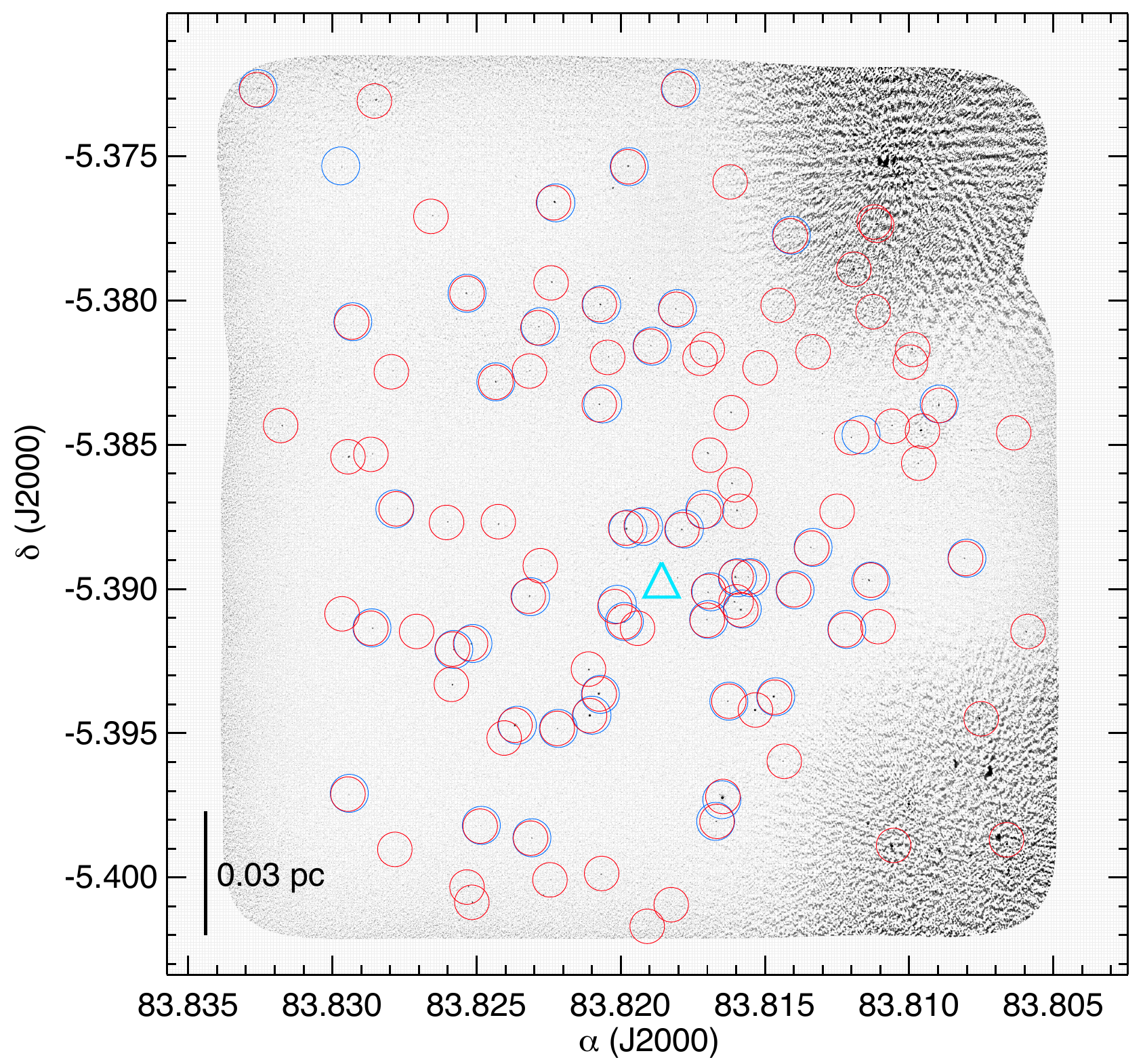}
\caption{Image of the central $1.5'\times 1.5'$ of the ONC, generated
  from 136 ALMA pointings at 350 GHz frequency.  A 100 k$\lambda$
  $uv$ cut has been employed, which substantially reduces the extended
  emission in the BN/KL and OMC1S regions (at the upper and lower right,
  respectively).  As described in Section \ref{sec:obs}, larger $uv$
  cuts can eliminate more extended emission, but at the expense of
  overall rms; the 100 k$\lambda$ cut is optimal.  Circles indicate
  known cluster members where the flux in this image is $\ge 4$ times
  the local rms noise.  Blue circles represent optically-identified
  proplyds, and red circles indicate near-IR-identified young stars.
  The position of $\theta^1$ Ori C is indicated with a cyan triangle.
\label{fig:map}}
\end{figure*}

In this paper we present a new 850 $\mu$m wavelength interferometric
survey of the Orion Nebula cluster (ONC) with the Atacama Large
Millimeter Array (ALMA).  The ONC is a young, embedded
stellar cluster composed of hundreds of stars spanning a broad mass
range.   The Trapezium region alone contains hundreds of stars within
a several arcminute radius, and pre-main-sequence
evolutionary models \citep[e.g.,][]{BARAFFE+15} fitted to spectroscopic
and/or photometric data indicate that most stars are less than 
approximately one million years old \citep[e.g.,][]{PROSSER+94,HILLENBRAND97}.
The high stellar density and strong UV irradiation from the Trapezium
stars ($\theta^1$ Ori C in particular) produce a substantially
different environment than the lower-density star-forming regions
described above.   We can therefore investigate here  correlations of
disk properties with stellar or environmental properties in this
richly clustered--and hence typical--region of star and planet formation. 


\section{Observations and Data Reduction \label{sec:obs}}
\label{sec:obs}
We mapped the central $1.5' \times 1.5'$ region of the ONC.  
The map is comprised of 136 mosaicked pointings.
The fields were observed at $\sim 350$ GHz frequency, corresponding
to a wavelength of 850 $\mu$m.  Observations were taken on 13
September 2016. We observed with four spectral windows centered at
343, 345, 355, and 358 GHz.  Two windows covered 2.0 GHz bandwidth,
and the other two covered 1.875 GHz.

The ALMA pipeline reduction included standard flux, passband, and gain
calibrations.  Sources used for calibration include J0510+1800 and
J0541-0211.  Because our mosaic covered a large area containing
many compact sources as well as extended emission, the pipeline
reduction was unable to complete the imaging process.  Adopting the
flux, passband, and gain calibrated data, we completed the imaging
process by hand.

Substantial parts of the bandpass contained spectral line emission,
which would contaminate the continuum if not taken into account.  We
flagged spectral channels containing strong emission in each of the
observed fields.  Fields including the BN/KL region or the
OMC1S region contain stronger line emission, and were more heavily
flagged.  After flagging, we split off a line-free continuum dataset.

We imaged the continuum data using the CASA task {\tt tclean}.  We
generated images with cell sizes of $0\rlap{.}''02$, and employed a
robust weighting parameter of 0.5.  We generated clean boxes using an
iterative process where we placed boxes on all objects detected above
5$\sigma$, generated a cleaned image, then examined the residuals for
additional detections.  We made images with both Hogbom and
multiscale (0, $0\rlap{.}''1$, and $0\rlap{.}''3$ scales) 
clean algorithms. While there is little difference in the
results, the multiscale clean produced slightly lower rms and we
therefore use products from that algorithm in our analysis.

We used the clean components of bright sources as a starting model for
self-calibration.  After applying the self-calibration solutions, the
rms in the final map was improved slightly (a few percent).  We
therefore use the self-calibrated data for our analysis.

Because we are
interested here in protoplanetary disks with expected sizes smaller
than a few hundred AU, we employ a $uv$ cut to filter out extended
emission from large-scale outflows or the background molecular cloud
\citep[as done in previous work on the ONC; e.g.,][]{FELLI+93a,EISNER+08}.
Note that the $uv$ cut is actually applied at the same time as
the inversion and cleaning described above.
Eliminating large-scale emission can improve the noise in the vicinity
of compact disks, but
eliminating data can also degrade sensitivity.  To determine the
optimal balance, we generated images with $uv$ cuts ranging from 0
(i.e., keeping all data) to 300 k$\lambda$.  

The optimal rms across the mosaic 
is achieved with a $uv$ cut of 100 k$\lambda$.   Our final image thus
uses this $uv$ cut, along with multi-scale cleaning and self-cal as
described above.
Finally, we applied a primary beam correction to the mosaic.  Our
final 850 $\mu$m continuum image is shown in Figure \ref{fig:map}.

The synthesized beam FWHM is $0\rlap{.}''09$.  At the distance to
Orion, $\sim 400$ pc
\citep[e.g.,][]{SANDSTROM+07,MENTEN+07,KRAUS+07,HIROTA+07,KOUNKEL+17},
the linear resolution is approximately 35 AU.  
The 100
k$\lambda$ $uv$ cut employed in our imaging corresponds to a 
spatial scale of $\sim 800$ AU.  This scale is substantially larger than any
of the disks imaged in our sample, and hence we have not resolved out
any disk emission with this $uv$ cut.

One of the 1.875 GHz-wide windows included CO(3-2) at a rest
frequency of 345.796 GHz, and another covered HCO$^+$(4-3) at a
rest frequency of 356.734 GHz.  We subtracted the continuum from
the line emission, and then imaged it using the same procedure as
outlined above.  We generated two cubes for each line: one covering
0 to 20 km s$^{-1}$ with 0.5 km s$^{-1}$ channels; and one spanning
-25 to 75 km s$^{-1}$ with 2 km s$^{-1}$ channels.  The
wider-bandwidth cube is intended to map outflows and jets, while the
narrower cube allows mapping of Keplerian motions in protoplanetary
disks.  As in previous work \citep{EISNER+16}, we detect
little line emission from compact disks.  While copious
emission is detected from larger-scale outflows, we defer analysis of
such phenomena to future work.

\epsscale{1.0}
\begin{figure*}[tbhp]
\plotone{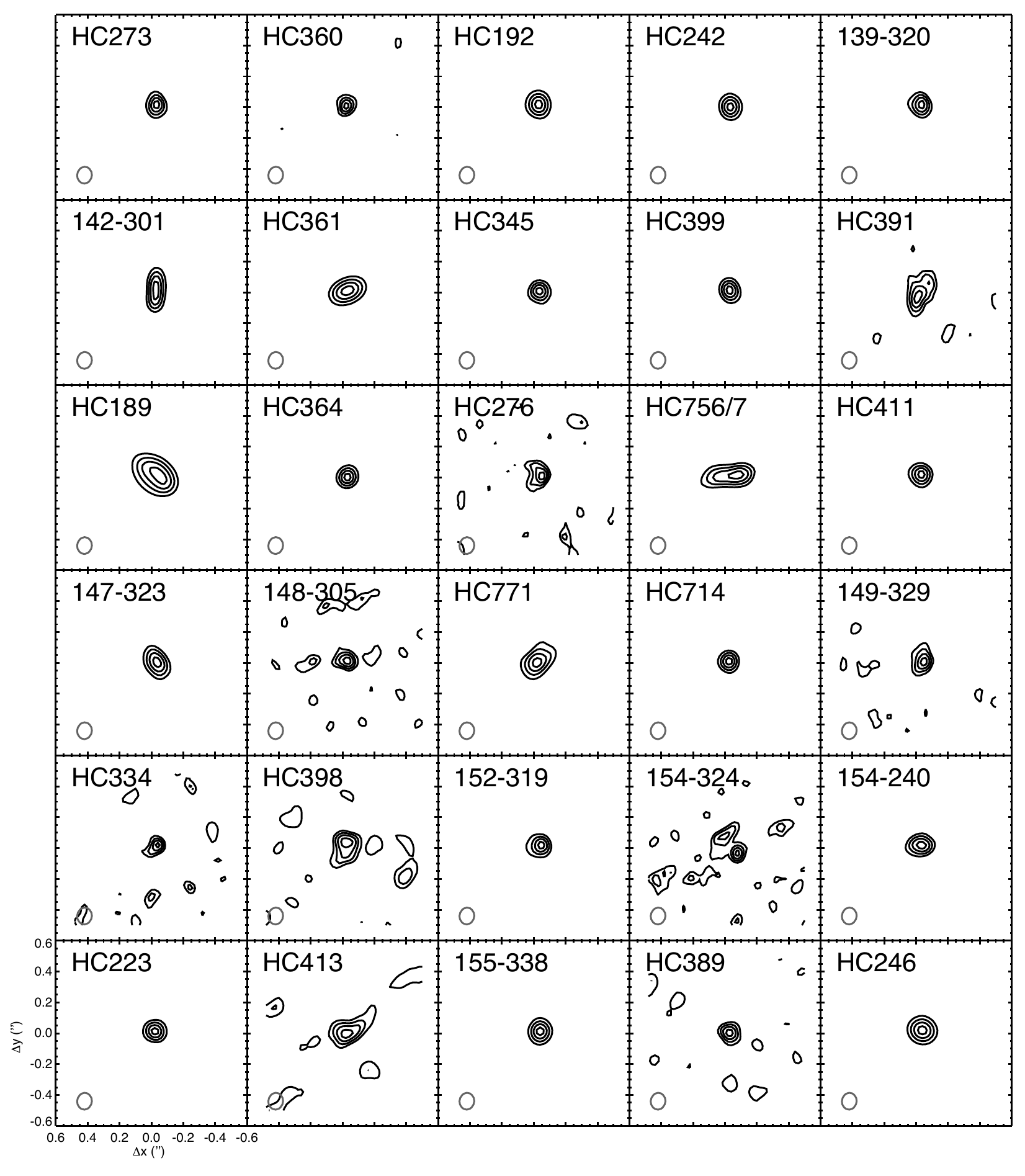}
\caption{Contour plots of $1\rlap{.}''2 \times 1\rlap{.}''2$ ($480
    \times 480$ AU) regions around each
  detected source (corresponding to the circles in Figure \ref{fig:map}).  Contours are drawn at
  30\%, 50\%, 70\%, and 90\% of the maximum flux in each region.
  The 50\% contour of the synthesized beam is indicated in the bottom left
  corner of each image.  Note that proplyds are indicated with
  six-digit IDs, while near-IR sources not detected as proplyds are
  labeled with ``HC'' and three digits.  HC 756 and HC 757 are a close
  binary pair, and are included in a single sub-image. 
\label{fig:detections}}
\end{figure*}

\begin{figure*}[tbhp]
\figurenum{\ref{fig:detections}}
\plotone{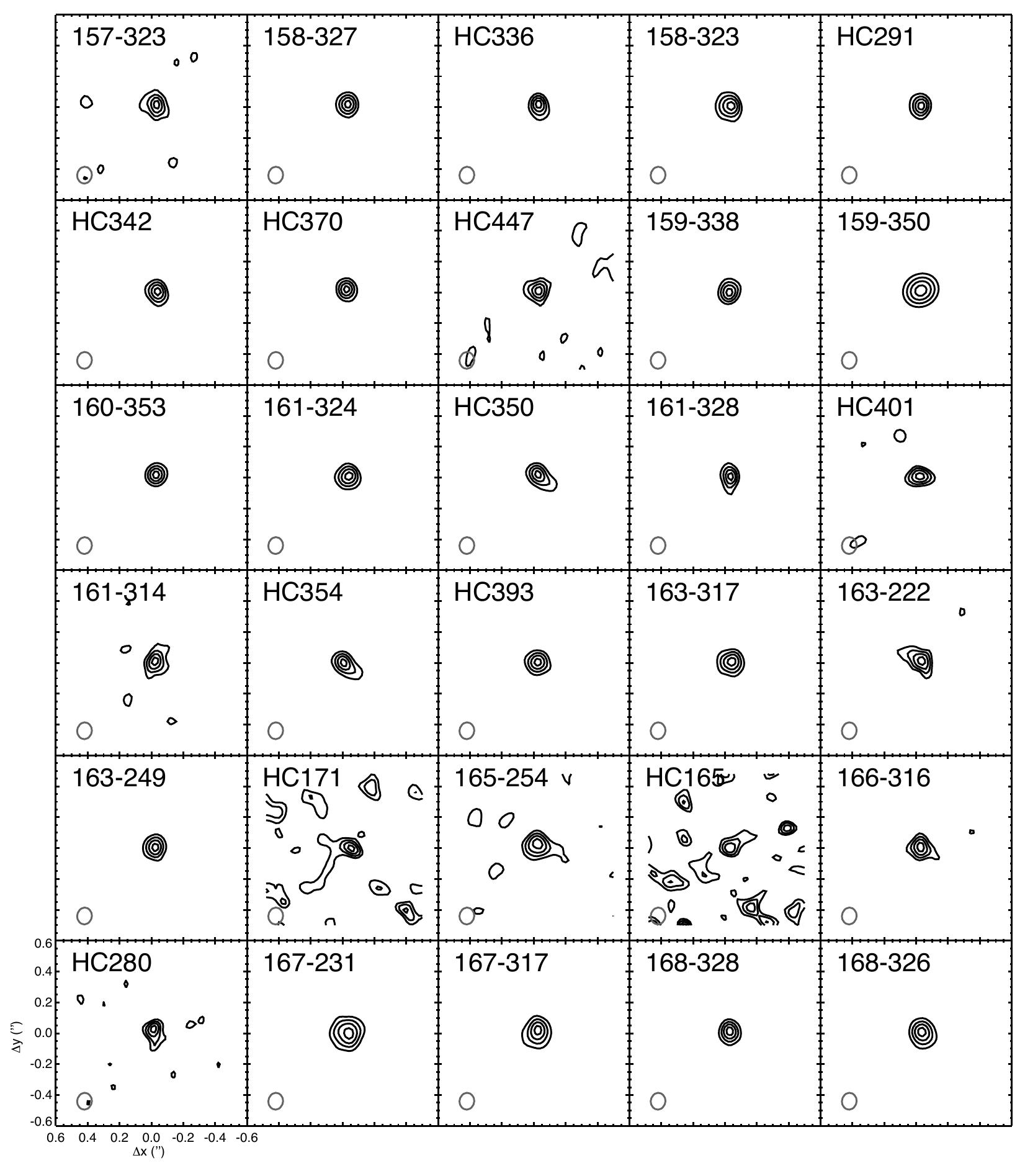}
\caption{continued.}
\end{figure*}

\begin{figure*}[tbhp]
\figurenum{\ref{fig:detections}}
\plotone{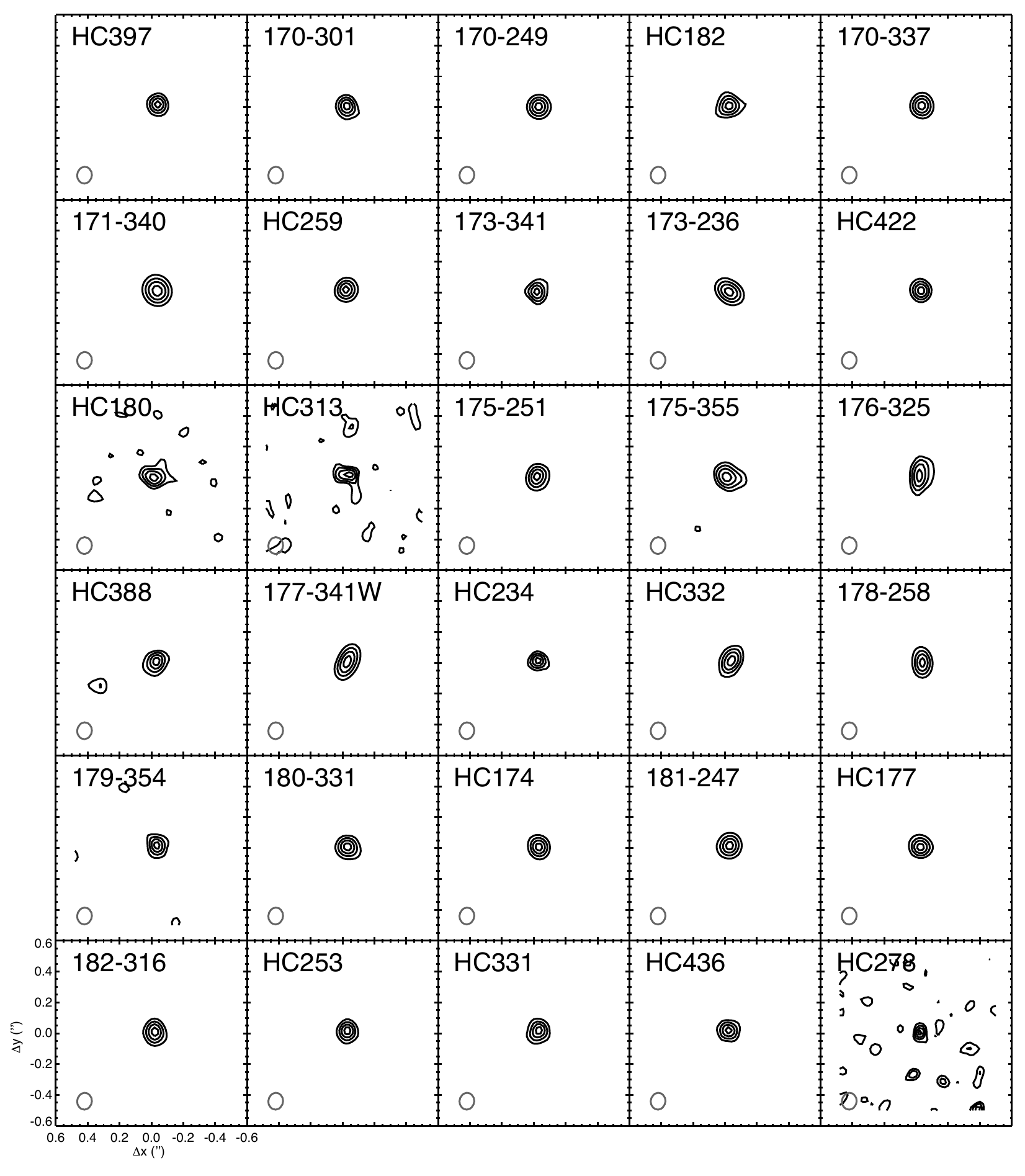}
\caption{continued.}
\end{figure*}

\begin{figure*}[tbhp]
\figurenum{\ref{fig:detections}}
\plotone{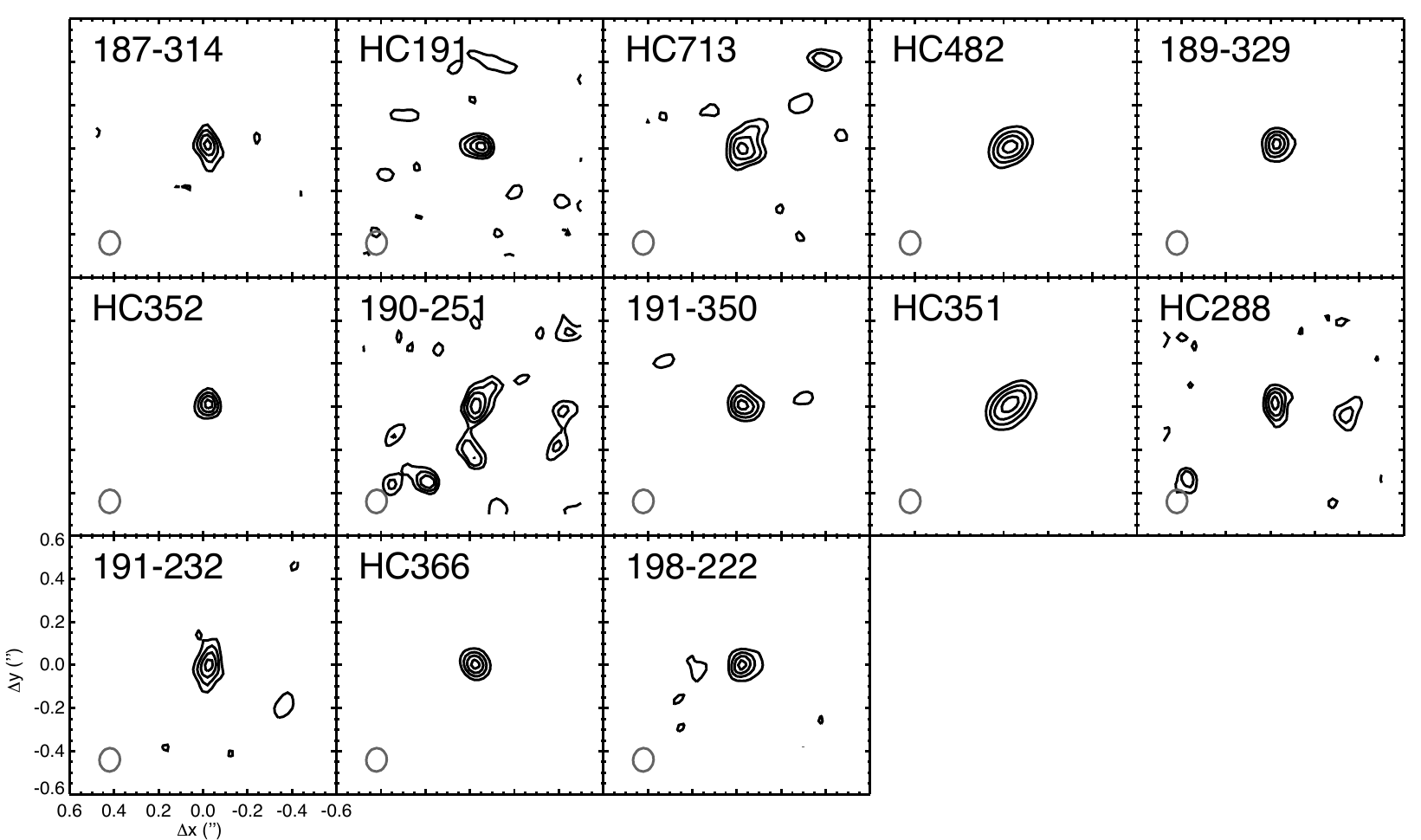}
\caption{continued.}
\end{figure*}

\section{Results}
We searched for 850 $\mu$m continuum emission towards the positions of
HST-detected proplyds \citep{RRS08} and near-IR detected sources
\citep{HC00,EISNER+16}\footnote{Most of the proplyds are also detected as
  near-IR objects.}.  
Given the positional uncertainties in the proplyd and
near-IR catalogs, we require optical/near-IR and sub-mm source
positions to coincide within $0\rlap{.}''5$.
We employed a detection threshold of
4$\sigma$ above the locally-determined noise level for each optical/near-IR
source.  This threshold ensures that $\ll 1$ detection is expected
from Gaussian noise fluctuations across the entire sample of known cluster
members. 

Because the noise in the map is higher in the vicinity
of bright, extended emission, we calculated a local noise level for
each source position.  In ``clean'' regions of the image, the rms is $\sim 0.1$
mJy, while it rises as high as 2.5 mJy in the BN/KL region.  We
verified that the mean level of sub-mm background around each detected object is
zero, even in the vicinity of bright extended emission (i.e., the
compact disk emission does not lie on top of extended emission).

Our ALMA mosaic includes 222 near-IR-identified young stellar objects,
and 61 proplyds (56 of
which are also seen in the near-IR).  We detected sub-mm continuum
emission towards 48 of the known proplyds, a detection rate of $\sim
80\%$.  The detection rate among the larger sample of near-IR targets
was somewhat lower, $\sim 50\%$, with sub-mm emission seen toward 102
near-IR-selected objects.  

In total, sub-mm continuum emission above the 4$\sigma$ level was seen
towards 104 cluster members (Table \ref{tab:detections}).  
Of these 104 cluster members, 48 are proplyds, and 102 are
near-IR-identified; i.e., two of the detected proplyds do not have
known near-IR counterparts.  49 of these objects were
detected in previous, less sensitive observations at mm or sub-mm
wavelengths \citep{MLL95,EC06,EISNER+08,MANN+14,EISNER+16}.
The remaining 55 objects are detected for the first time at these
wavelengths. Sub-images
toward each detected object are shown in Figure \ref{fig:detections}.

In Table \ref{tab:nondet}, we list the cluster members not detected
above 4$\sigma$ in our ALMA map.  For each source we give the
optical/IR source position, and the 4$\sigma$ upper limit computed
from the local rms noise level.  

\epsscale{0.9}
\begin{figure*}
\plotone{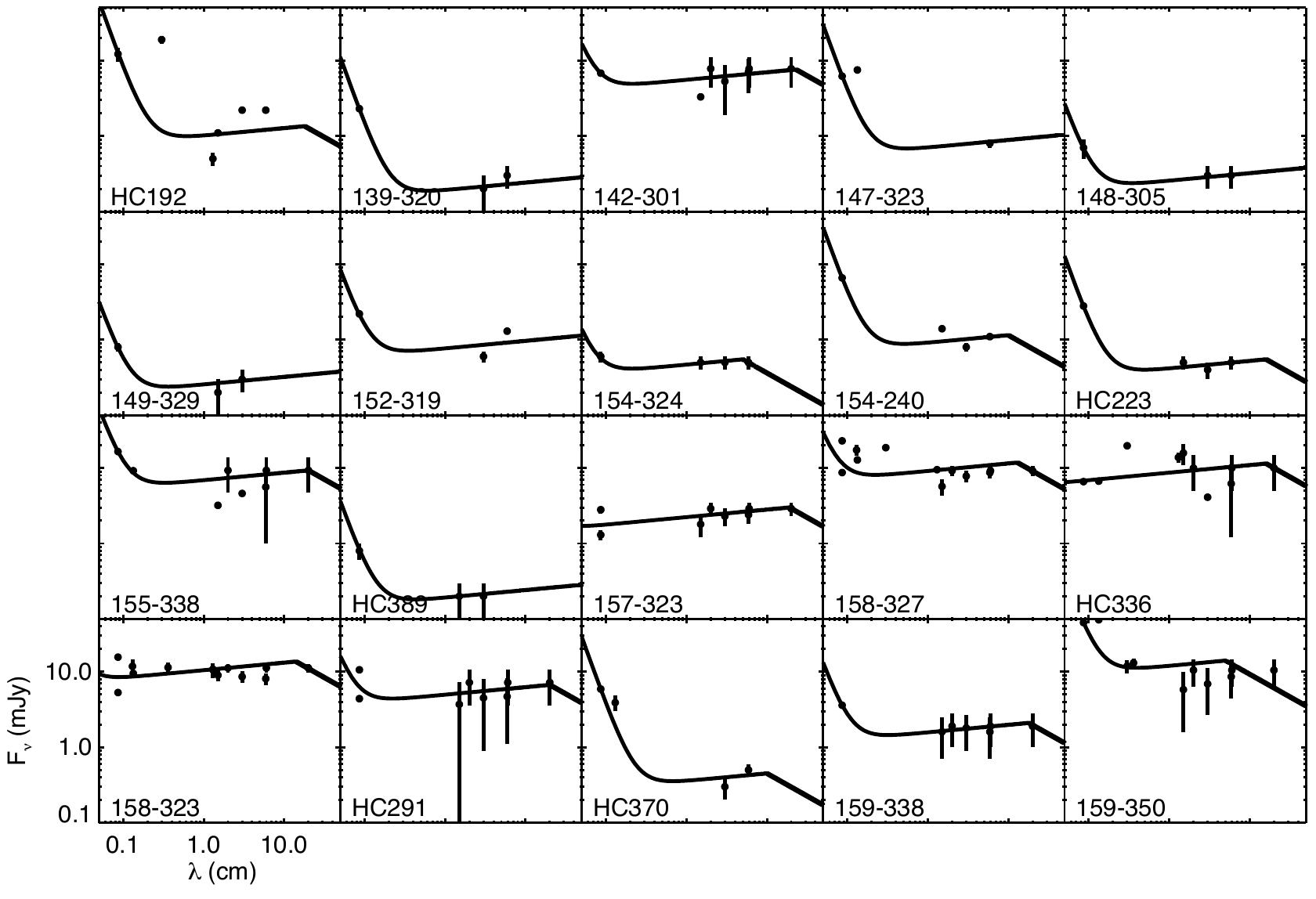}
\caption{Long-wavelength fluxes for ONC cluster members detected in
  our ALMA observations (points with 1$\sigma$ error bars), and
  best-fit models including free-free and dust emission
  (curves). Error bars are smaller than the plotted symbols in some cases.
\label{fig:ff}}
\end{figure*}

\begin{figure*}
\figurenum{\ref{fig:ff}}
\plotone{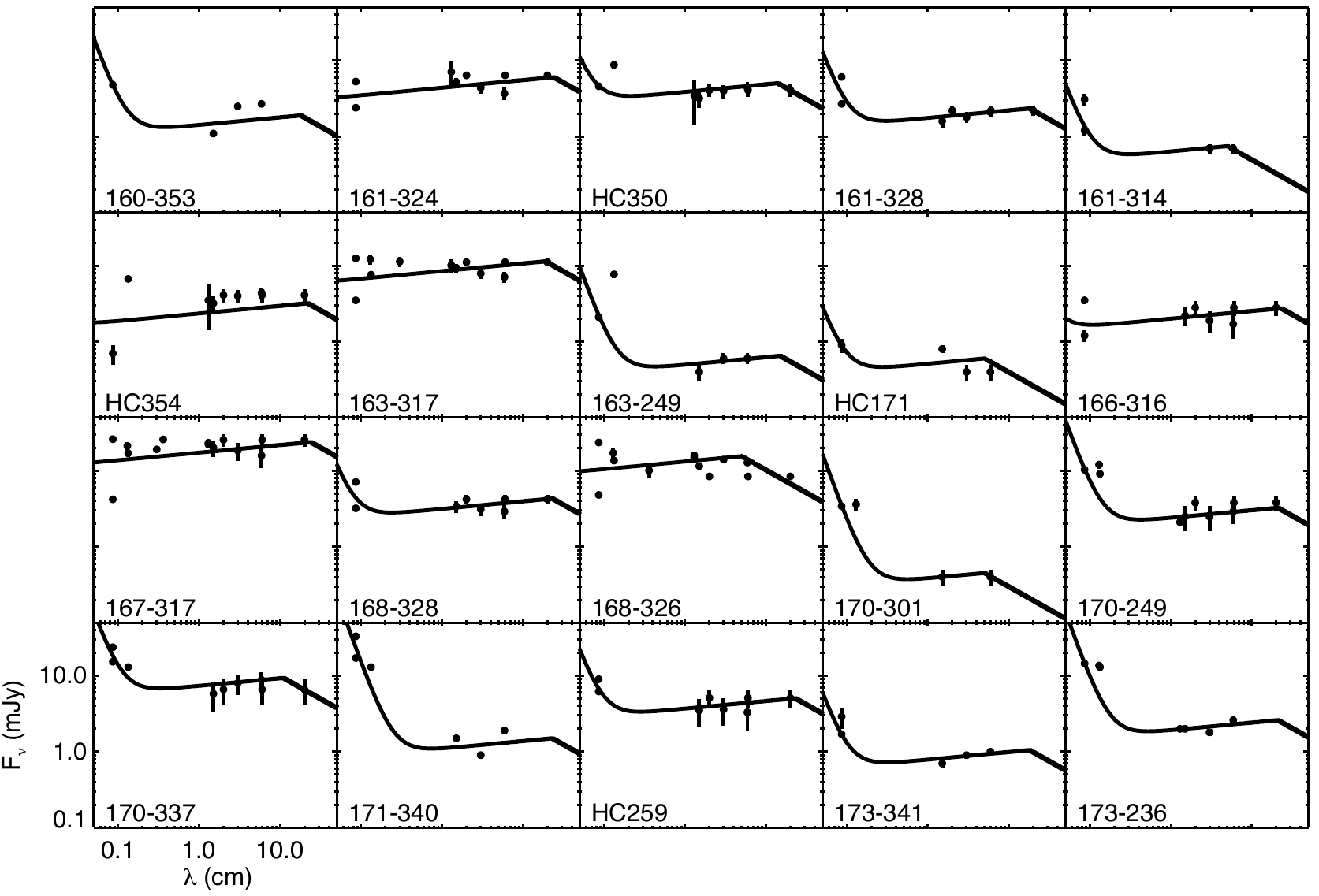}
\caption{continued.}
\end{figure*}

\begin{figure*}
\figurenum{\ref{fig:ff}}
\plotone{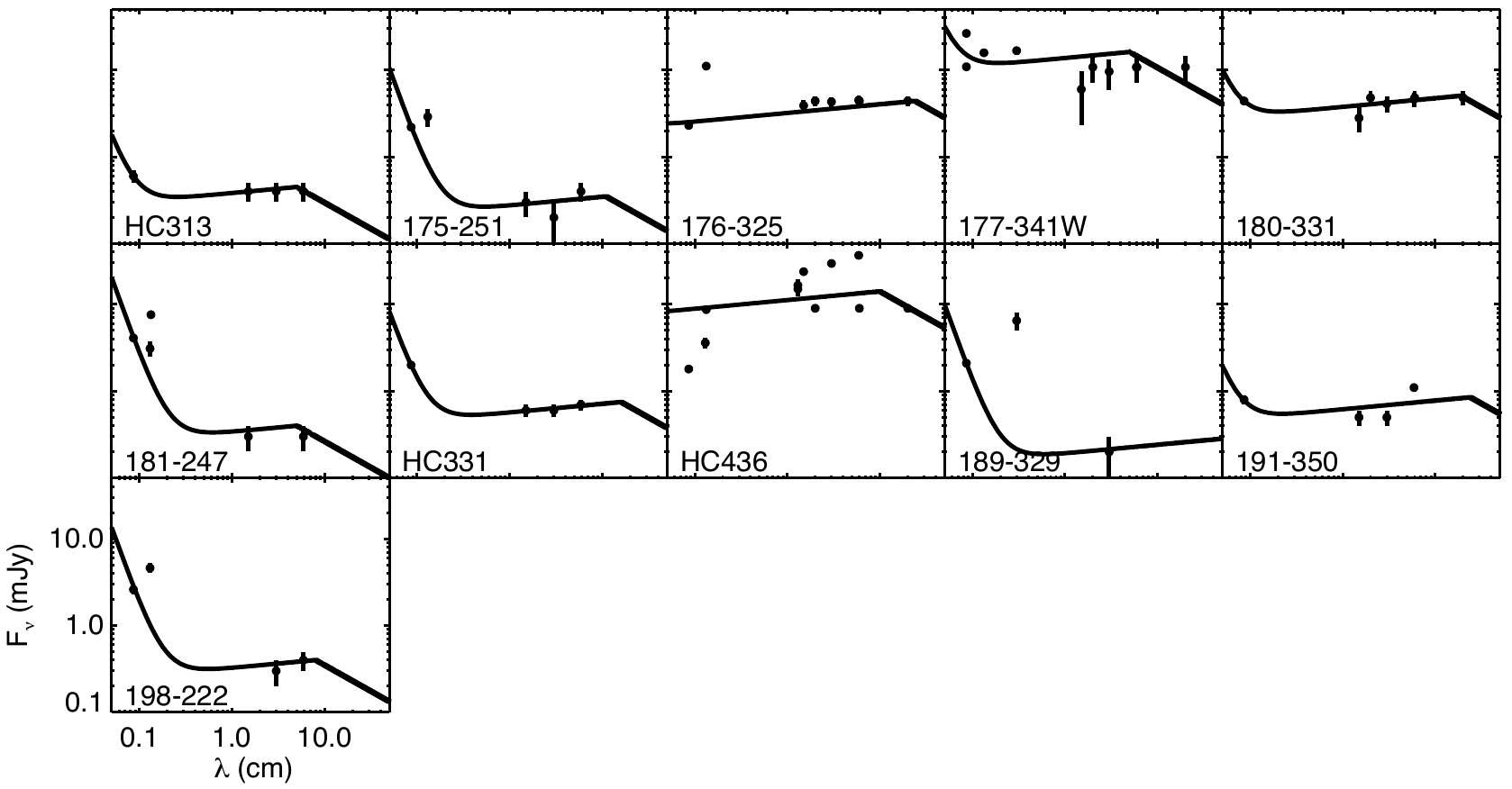}
\caption{continued.}
\end{figure*}

Given the strong ionization field near the Trapezium stars, gas in
circumstellar disks or outflows can emit free-free emission.
Contributions of sub-millimeter-wavelength free-free emission must be
quantified to accurately determine the flux arising from
dusty disk matter.  The spectrum of optically thin free-free emission
is relatively flat compared to that of dust emission, and so
observations at cm wavelengths can be used to constrain free-free
emission \citep[e.g.,][]{EISNER+08,SHEEHAN+16}. 

We used previous observations of the ONC at cm wavelengths
\citep{FELLI+93a,FELLI+93b,ZAPATA+04,FORBICH+07,FORBRICH+16,SHEEHAN+16} 
to search for free-free emission from our sample.  These surveys used
the VLA, and have angular resolution comparable to the resolution in
our ALMA observations.  The more recent
of these surveys have the sensitivity required to constrain free-free emission
at the same flux levels seen in our ALMA map.  52 of the sources detected in
our ALMA observations are also seen in these previous observations.

For the 52 targets detected previously at cm wavelengths, 
we constructed a model that includes optically
thick free-free emission past a turnover wavelength (typically $>1$--10
cm),  thin free-free
emission short-ward of that turnover, and optically-thin dust with
$\beta=1$ \citep[as in][]{EISNER+16}. 
Free parameters in the model
are the turnover wavelength of free-free emission, and the free-free
and dust fluxes at the observed ALMA wavelength.
Best-fit models are shown in
Figure \ref{fig:ff}. 

There are a few objects in Figure \ref{fig:ff} for which the
best-fit models do not fully capture the scatter in the photometric
data. Examples include HC 192 and 198-222.  The scatter in the
photometry for these sources probably reflects intrinsic variability
\citep[which has been seen
previously; e.g., ][]{FELLI+93b,ZAPATA+04,KOUNKEL+14,RIVILLA+15,SHEEHAN+16}.
If the variable emission arises from dust and ionized gas, then our
use of all photometry in the fits ensures the scatter in the data is reflected
in the error bars listed in Table \ref{tab:detections}.   However our
model does not include potential contributions from gyrosynchroton
emission, which may cause (some of) the large-amplitude radio
variability in a small fraction of ONC members \citep[e.g.,][]{SHEEHAN+16}. 

The remaining objects detected in our ALMA mosaic are not seen in
cm-wavelength observations.  These non-detections imply that
the cm-wavelength flux, and hence the potential free-free contribution
at mm wavelengths, is $\la 0.03$ mJy for these sources.  Since the
observed ALMA fluxes are all $\ge 0.5$ mJy, we can be confident that
we have detected dust emission.  The $\la 0.03$ mJy uncertainties
resulting from potential low-level free-free emission are much smaller than
the uncertainties in the mm-wavelength flux measurements.

\begin{figure*}[tbp]
\plottwo{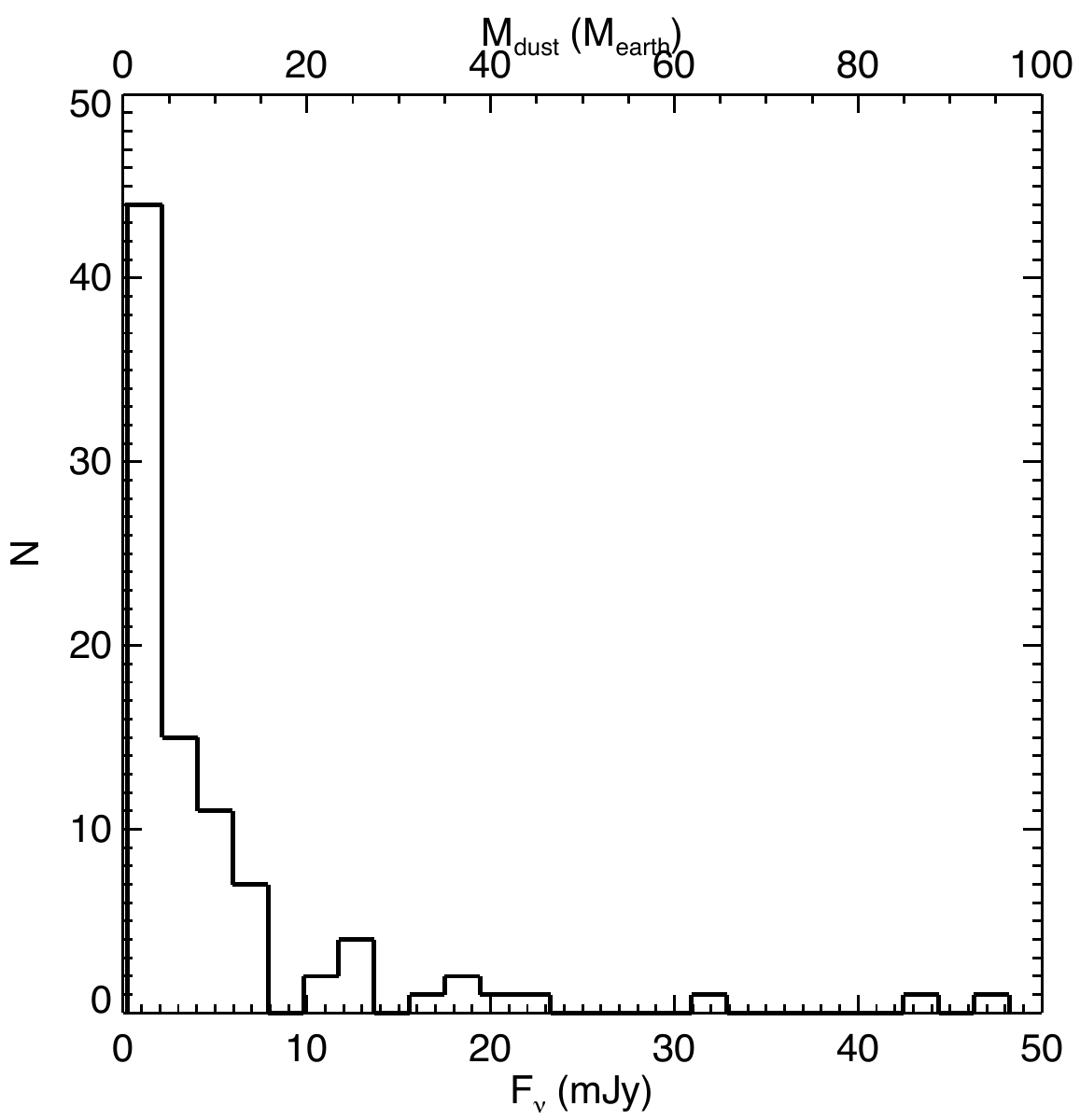}{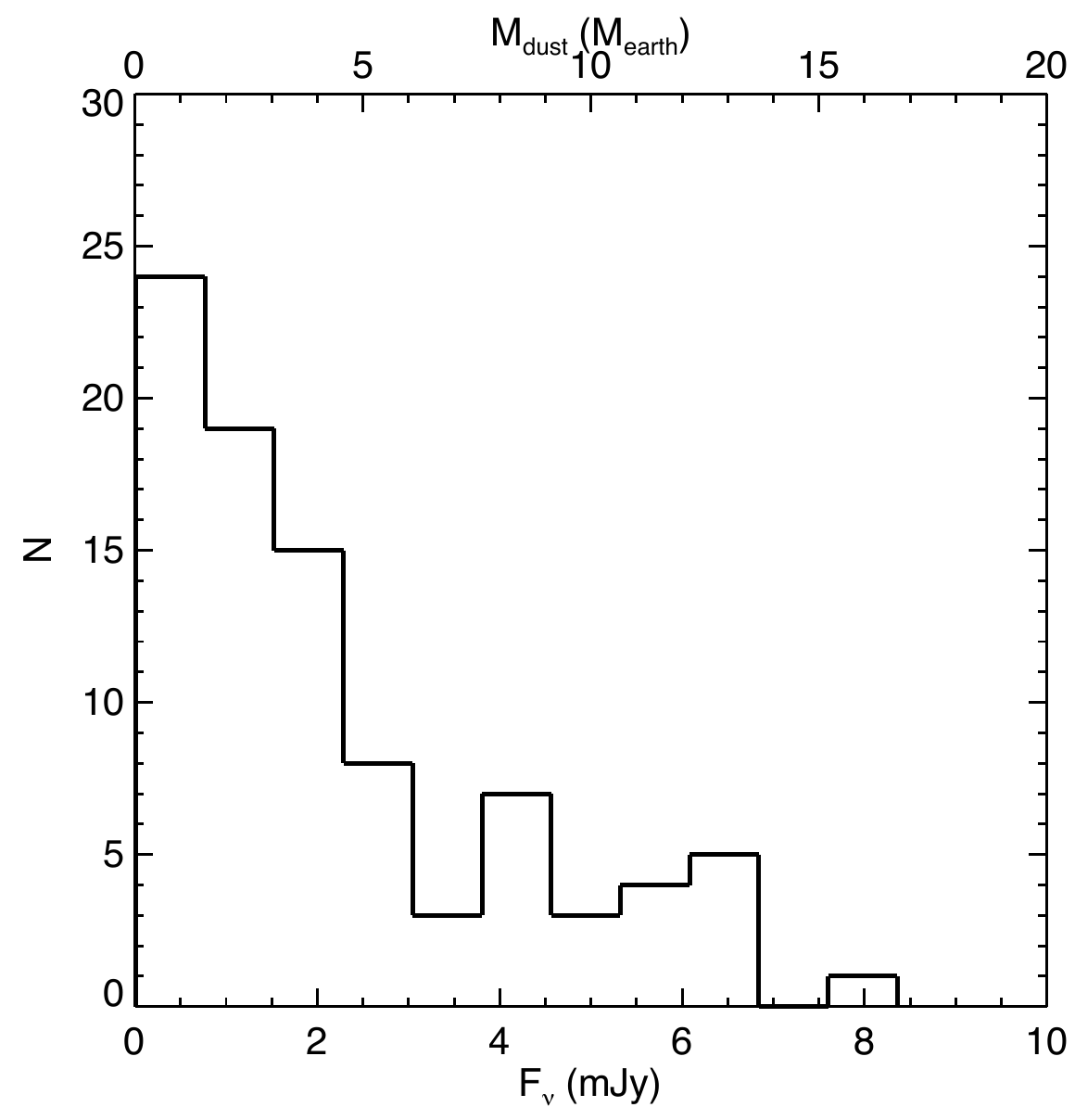}
\caption{({\it left:}) The distribution of measured 350 GHz ($\lambda$850 $\mu$m)
  continuum fluxes for  detected objects, after subtraction of
  contributions from free-free emission.  We also include an
  additional axis label indicating the
  masses corresponding to these fluxes, calculated using Equation
  \ref{eq:dustmass}.  The assumptions inherent in Equation
  \ref{eq:dustmass} are almost certainly incorrect for this sample
  (see Section \ref{sec:mdist}),
  and the masses are listed here to provide a point of reference
  only.  ({\it right:}) The distribution of fluxes $<10$ mJy for
  detected objects.
\label{fig:fdist}}
\end{figure*}

We plot a histogram of measured fluxes, corrected for free-free
contributions, in Figure \ref{fig:fdist}.  Free-free corrected fluxes
are also listed in Table \ref{tab:detections}.  Figure
\ref{fig:fdist} includes only dusty disks; i.e., sources for
which the sub-mm flux can be attributed purely to free-free emission
are excluded.  The distribution of
corrected fluxes is strongly peaked around $\sim 0.5$ mJy.
This flux is close to the 4$\sigma$ detection limit, which is $\ga
0.4$ mJy across the
mosaic.  

\epsscale{0.9}
\begin{figure}[tbhp]
\plotone{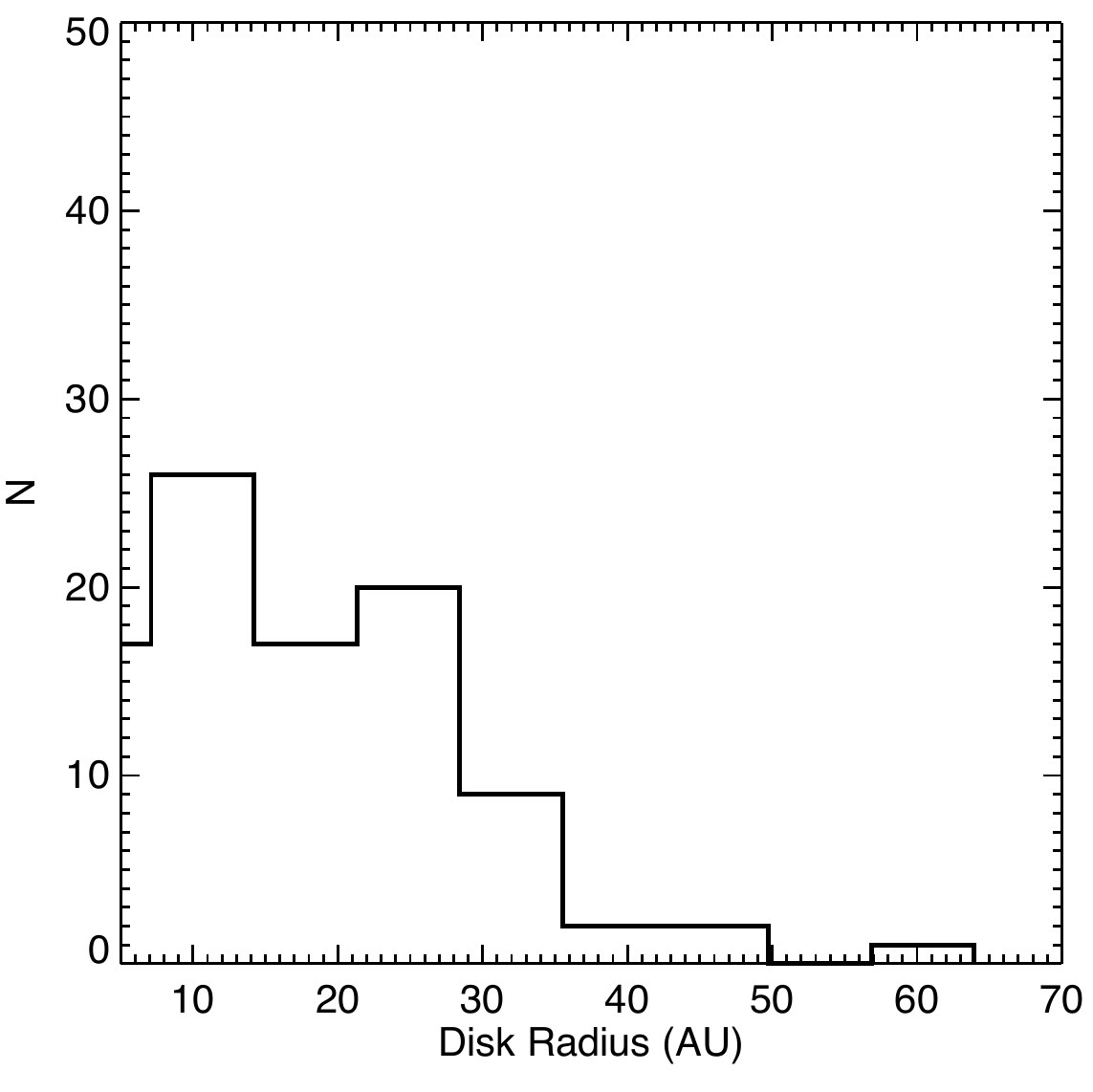}
\caption{Distribution of disk sizes for our sample of dusty disks.  
These radii represent the HWHM major axes of Gaussian fits to images
of each detected
object (shown in Figure \ref{fig:detections}).  The lowest-radius bin
includes the 15 unresolved objects, which have radii $\la 5$ AU. 
\label{fig:sizes}}
\end{figure}

The  $\approx 0\rlap{.}''09$ (35 AU) angular resolution of our
observations is
sufficient to constrain sizes of detected disks.  To measure sizes, we
fitted a 2D elliptical Gaussian to each of the sub-images shown in
Figure \ref{fig:detections}.  We used Gaussian deconvolution to remove
the effect of the beam.  Even though the beam has a FWHM of $\sim 40$
AU, objects with intrinsic FWHM $\ga 10$ AU lead to sources that are
resolved, since a 10 AU source convolved with a 40 AU beam yields an
``observed'' image with FWHM larger than 40 AU.  Thus, objects in our
sample where the fitted size is comparable to that of the beam are
considered to have intrinsic FWHM $<10$ AU (i.e., HWHM $<5$ AU).  

Beam-deconvolved disk radii (HWHM of the major axis of 2D Gaussians)
are listed in Table \ref{tab:detections}.
Of the 104 detected sources, all but 15 are spatially resolved.
In Figure \ref{fig:sizes}, we show a histogram of disk sizes for dusty
disks.  These are sources with non-zero dust emission (i.e., excluding
those sources in Table \ref{tab:detections} for which 100\% of the
flux is attributed to free-free emission).  The distribution is peaked around disk
radii of $\sim 10$--30 AU, with few disks of radii larger than 35 AU.

 \section{Discussion}

\subsection{Disk Detection Rates \label{sec:rates}}
The detection rate among optical and near-IR-selected cluster members
that lie within the ALMA mosaic is $\sim 50\%$, lower than the
detection rate of $\sim 80\%$ for only the  HST-identified proplyds.
A similar discrepancy has been seen in
previous work, albeit with smaller numbers of detections
\citep[e.g.,][]{EISNER+16}.  

Since we are interested primarily in sources with significant dust emission,
rather than objects where free-free emission is found
above the detection threshold of our observations, we first examine
the detection rate of dusty disks.  Of the 104 detected objects listed
in Table \ref{tab:detections}, 12 are consistent with 100\% of the
sub-mm continuum flux arising from free-free
emission.  Of these, 8 are HST-identified proplyds.  Removing these
free-free-dominated sources from the sample, the detection rate of
dusty disks in the proplyd sample is $\sim 65\%$.  In the larger
sample of cluster members, dusty disks are
detected around $\sim 45\%$ of sources.  

Given the sample size, the
discrepancy between the proplyd sample and the larger near-IR sample
(or, by extension, the non-proplyd component of the near-IR sample) is
marginally statistically significant.  $65 \pm 10\%$ of proplyds have dusty disks,
while $45 \pm 7\%$ of all cluster members in the region have dusty disks.

One possible explanation for the higher detection rate among the
proplyd sample is that some of the near-IR sample are not {\it bona
  fide} ONC cluster members.  Since proplyds typically have spatially
resolved, disk-like morphologies, one does not expect contamination of
that sample by background objects.  While some background or
foreground source contamination of the near-IR-selected sample is
possible, \citet{HC00} estimate that $<5\%$ of objects are contaminants.

A similar, alternative explanation, is that the
near-IR sample may simply be less likely to have massive disks.  The
imaging-selected proplyd sample is likely to contain large, and hence
massive disks.  While $H-K$ excess from the near-IR-selected sample
suggests the presence of disks around nearly all objects \citep{HC00},
there is no {\it a priori} reason to expect these disks to be large.

In any event, the detection of dusty disks around $\ga 45\%$ of ONC
cluster members
suggests that we are probing close to the peak of the disk mass
distribution with the sensitivity achieved in our observations.
Most detected sources have 850 $\mu$m continuum fluxes around
0.5 mJy, which is close to our detection threshold.  Thus, we expect
the disk detection rate to increase with any increase in the
sensitivity of future observations.  

\subsection{Disk Size Distribution \label{sec:sizedist}}
The disks in the ONC appear to be compact in comparison with disks in
other regions.  Disk sizes have been measured in several low-density
star forming regions, some with ages similar to the $\sim 1$ Myr age
of the ONC, and some with substantially older ages.  We consider
Taurus, Ophiuchus, Lupus, Cham I, and Upper Sco, which have
inferred ages of 1--2 Myr, 1--2 Myr, 1--3 Myr, 2--3 Myr, and 10 Myr,
respectively \citep[see][and references therein]{PASCUCCI+16}.

\epsscale{1.0}
\begin{figure}[tbhp]
\plotone{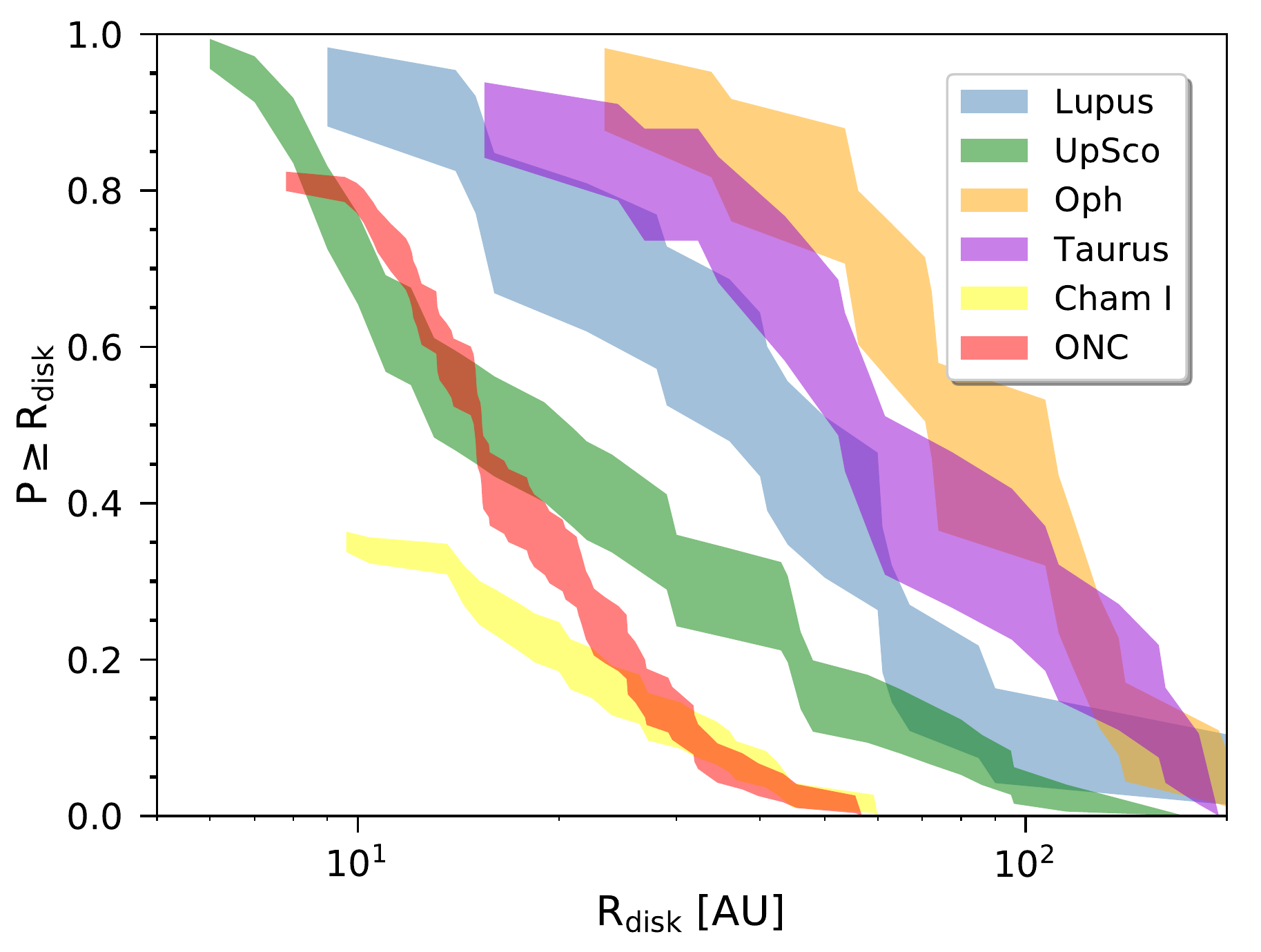}
\caption{Disk size cumulative distribution for the ONC compared
 with previous work in Taurus and Ophiucus \citep{TRIPATHI+17}, Lupus
 \citep{TAZZARI+17}, Cham I \citep{PASCUCCI+16}, and Upper  Sco
\citep{BARENFELD+17}.  The distributions and the 1$\sigma$ confidence
intervals (shaded regions) were calculated using the Kaplan-Meier
estimator in order to account for upper limits in the samples
\citep[e.g.,][]{LAVALLEY+92}.
Since unresolved disks are included as censored data, the
distributions do not always reach unity.
While disk size is quantified via
Gaussian HWHM for the ONC, Cham I, Taurus, and Oph, in Lupus and
Upper Sco the radius is computed as the exponential cutoff radius of
a power-law disk.
\label{fig:sizecdf}}
\end{figure}

In Figure \ref{fig:sizecdf} we show the cumulative probability density
funtion for disk size in the ONC and other regions.  For the ONC, we
compute disk radius as a Gaussian HWHM.  The same quantity is
calculated for Cham I \citep{PASCUCCI+16}, Taurus, and Ophiuchus 
\citep{TRIPATHI+17}.  In Lupus and Upper Sco, the
inferred radius is the exponential cutoff radius of a power-law disk
\citep{TAZZARI+17,BARENFELD+17}.   While this quantity may differ
slightly from disk HWHM, the difference is likely smaller
than a factor of two \citep[e.g.,][]{EISNER+04}.  The discrepancies in
size distributions between the ONC and other regions are considerably
larger than this uncertainty.
Furthermore, while different surveys
obtained different sensitivities and angular resolutions, all were
capable of detecting the largest disks, which are clearly lacking in
the ONC.

In the younger regions, Taurus, Ophiuchus, and Lupus, disk sizes are
substantially larger than the distribution observed in the ONC.  In
the older Cham I and Upper Sco regions, many compact disks are
observed, similar to the disks seen in the ONC.  However, Upper Sco
contains a number of disks with inferred sizes substantially larger
than anything seen in the ONC sample.

\begin{figure}[tbhp]
\plotone{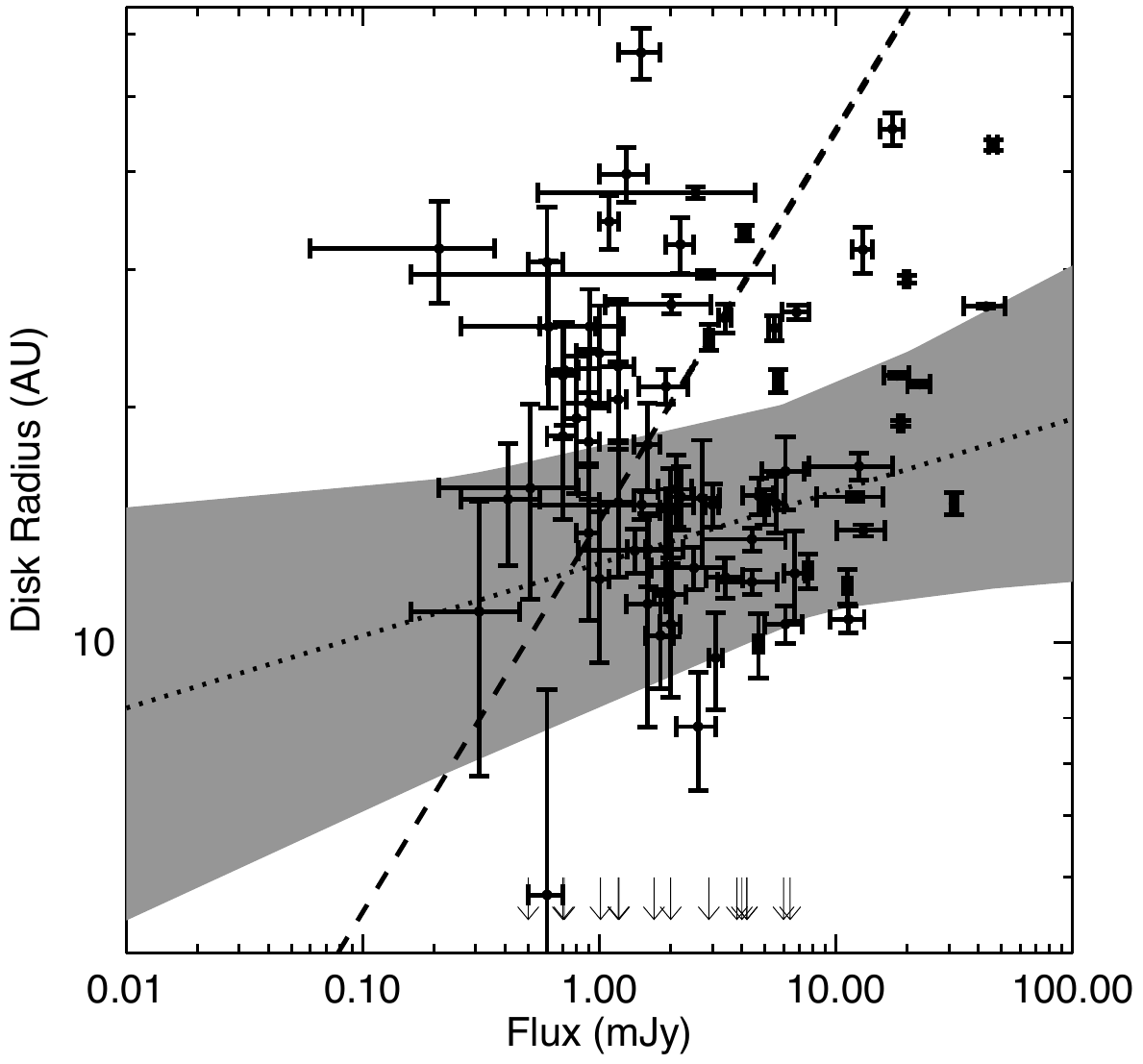}
\caption{Fitted disk radius (HWHM) versus sub-mm flux for the dusty
  disks detected in our ONC map (see Table \ref{tab:detections}). 
  Unresolved sources are indicated as upper
  limits.     We performed a linear regression to these data, including
  uncertainties and upper limits, using the method of
  \citet{KELLY07}.  The median of the posterior distribution is
  indicated with a dotted line, and the 1$\sigma$ confidence interval
  is shown as a gray shaded region.  The corellation is only
  marginally statistically significant in this case: log (HWHM/AU) $= (1.1
  \pm 0.2) + (0.09 \pm 0.07)$ log (Flux/mJy). The dashed line indicates
  the relationship determined by \citet{TRIPATHI+17} in low-density
  star-forming regions:  HWHM $= 160 L_{\rm mm}^{0.5}$,  
  where $L_{\rm mm}$ is the 850 $\mu$m flux in Jy, scaled to a distance of
  140 pc.  This trend lies
  outside of the 1$\sigma$ region of the posterior distribution of the
  ONC data regression.  To use another metric, the trend from
  \citet{TRIPATHI+17} has more than four times larger $\chi^2$ residuals than
  the median of the posterior distribution from the regression to the
  ONC data.
\label{fig:sizeflux}}
\end{figure}

In Figure \ref{fig:sizeflux} we plot measured disk radius versus sub-mm flux for
our ALMA ONC sample.  
We applied linear regression
using the Bayesian analysis described in \citet{KELLY07}.  This
technique takes into account heteroscedastic error bars on both
parameters of interest, and uses available upper limits to constrain
the regression.  

While no statistically significant trend is seen, we
do exclude any steep dependence of disk size on dust mass.
Thus, allowed slopes for the ONC sample are shallower than the relation
derived for lower-density regions \citep{TRIPATHI+17,TAZZARI+17}.   
However it is worth noting that the flux-size data in lower-density regions
span a larger range of sub-mm flux and disk size, and
hence have a better lever with which to determine the slope.

\epsscale{1.0}

The compact size distribution of ONC disks means that the sub-mm
fluxes in Table \ref{tab:detections} correspond to
dust emission from potentially planet-forming zones.  Indeed, it
may be that the $\sim 30$ AU radius of our own solar system is related
to the original size of the protosolar nebula set by the external
environment.  

We can obtain a rough estimate of the scale set by
photoevaporation by considering the gravitational radius, $2 G M_{\ast}
/c_{\rm s}^2$.  At larger stellocentric radii, the sound speed exceeds
the escape velocity, and photoevaporation causes mass loss.
Assuming a 0.3 M$_{\odot}$ star \citep[typical for the
ONC; e.g.,][]{HILLENBRAND97}, and a sound speed of 5 km s$^{-1}$
\citep[e.g.,][]{HO99}, the
gravitational radius is $\sim 20$ AU, similar to the disk radii
inferred for our sample.

Many of the proplyds in our sample have disk radii measured in the
optical with HST \citep{VA05}.  The optical source sizes are typically
larger than the sub-mm emission by a factor of a few.  Most optical
measurements of proplyds trace ionization fronts, and hence we expect
them to be larger than dense, dusty disks traced by sub-mm emission.
However even silhouette disks are larger in the optical, presumably
because even tenuous columns of dust can become optically thick at
optical wavelengths. In future work, we intend to model all available,
multi-wavelength, spatially resolved images simultaneously, to better
constrain disk sizes and dust properties (see Section \ref{sec:mdist}
for further discussion).

\begin{figure*}[tbp]
\plottwo{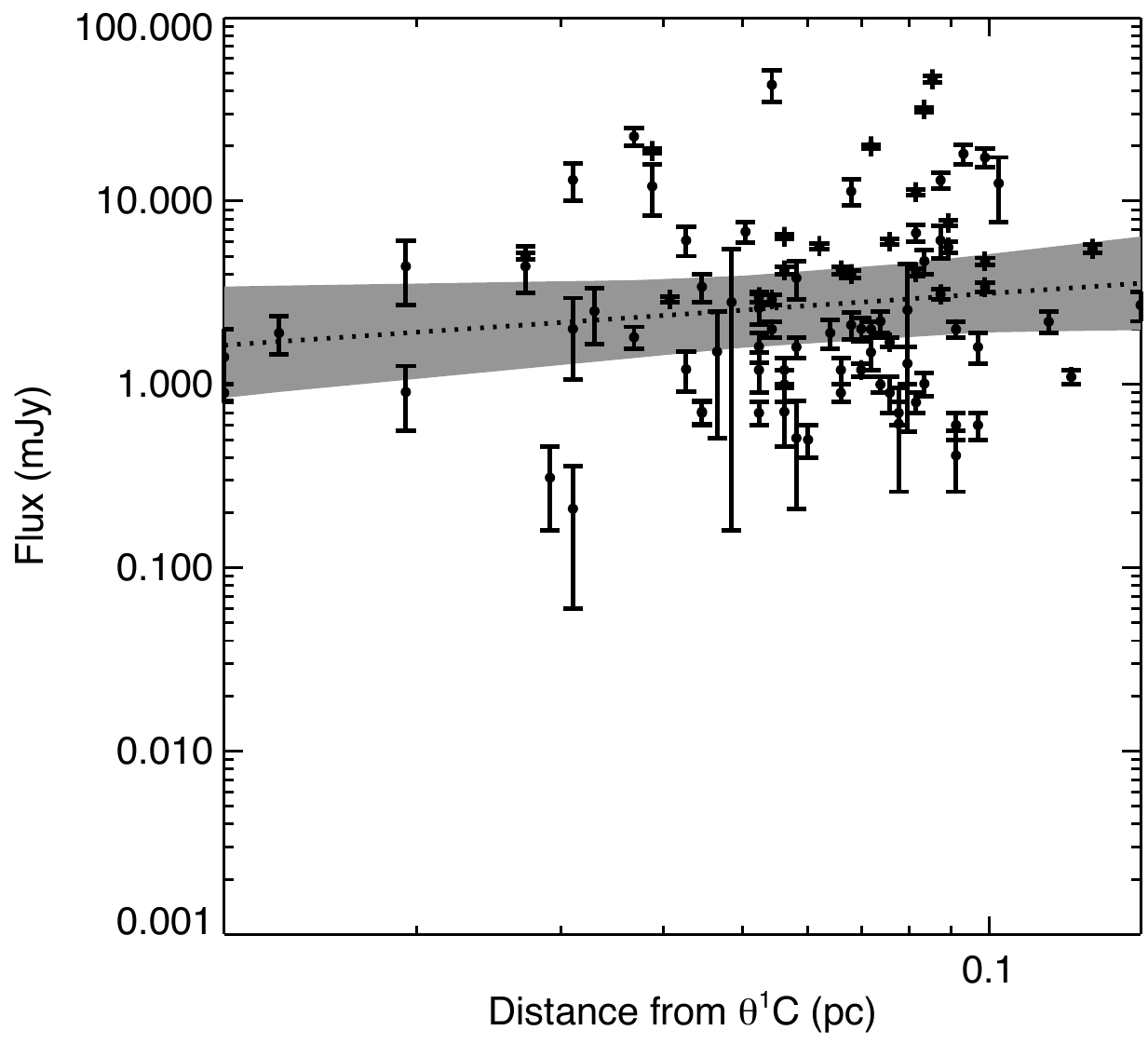}{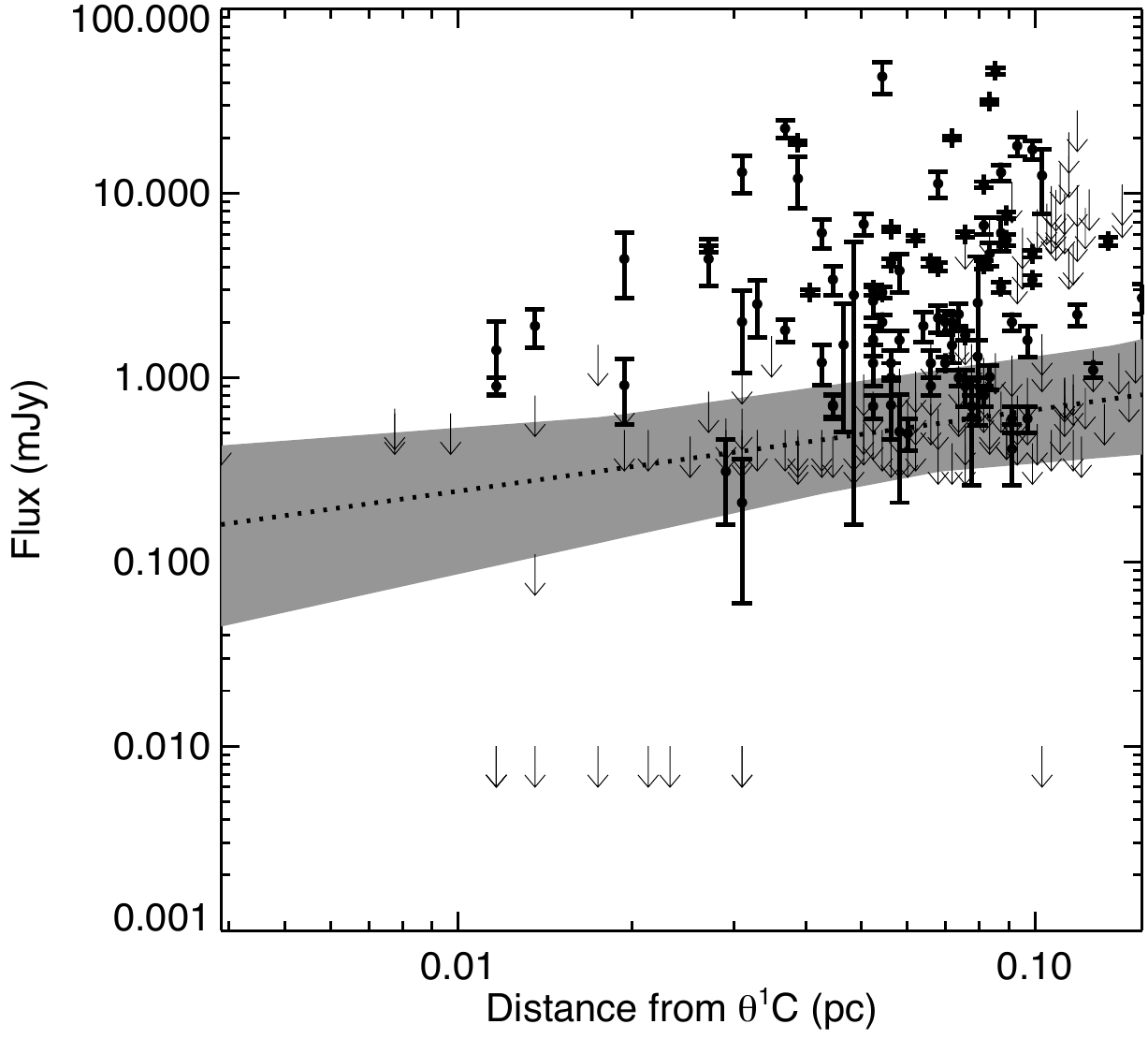}
\caption{({\it left}:)Sub-mm flux for dusty disks versus projected distance from
  $\theta^1$ Ori C, along with the best-fit linear regression and
  1$\sigma$ confidence region.   We see a paucity of disks with higher
  sub-mm fluxes within 0.03 pc of $\theta^1$ Ori C \citep[as seen in
  previous studies;][]{MANN+14,EISNER+16}.   Flux also
  correlates with distance to the Trapezium over the entire range of
  cluster radii probed here. The regression suggests log (Flux/mJy) = $(0.80
  \pm 0.45) + (0.30 \pm 0.25)$ log (Distance/pc).
  ({\it right:}) The same plot, but
  including all cluster members in the region, both detections and
  non-detections (either from Table \ref{tab:nondet} or
  free-free-dominated sources in Table \ref{tab:detections}). 
  We treat these
  non-detections as censored data when performing the regression.  
  The fitted trend is log (Flux/mJy) = $(0.27 \pm 0.25) + (0.45 \pm 0.20)$ log
  (Distance/pc),  consistent with (but more significant than) the
  trend in the left panel. 
\label{fig:clusterrad}}
\end{figure*}

\begin{figure}[tbhp]
\plotone{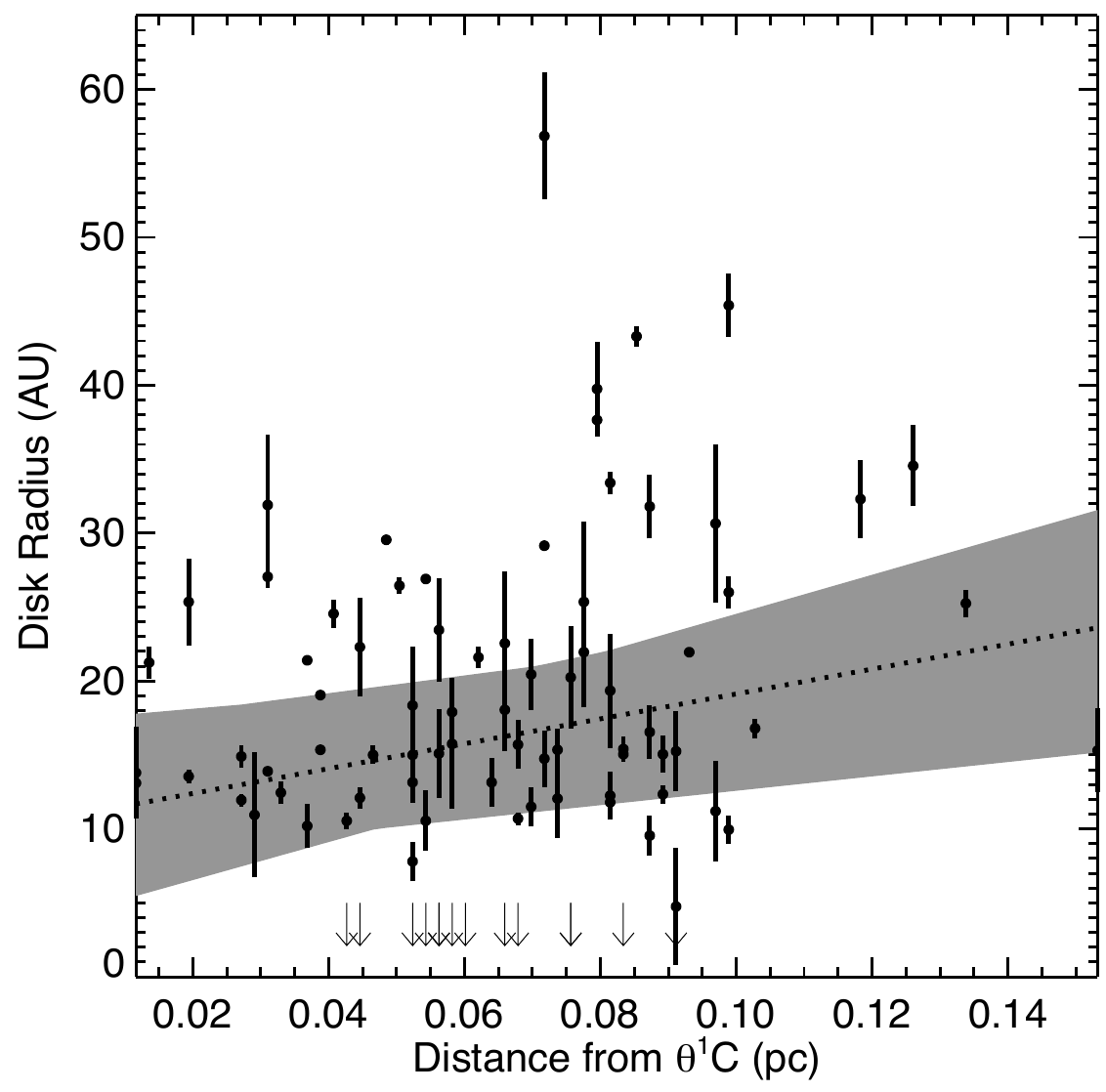}
\caption{Radii (HWHM) of dusty disks versus projected distance from
  $\theta^1$ Ori C.  
  We see a trend of size with cluster radius, (Radius/AU) = $(11 \pm 7) +
  (83 \pm 50)$ (Distance/pc).
\label{fig:sclusterrad}}
\end{figure}

\subsection{Disk Photoevaporation  \label{sec:radius}}
The lack of disks with radii larger than 50 AU in our ONC
sample is striking in comparison with low-density star-forming
regions.  We suggest
that this difference is due to the rich cluster environment in the
ONC.  Since external evaporation tends to strip outer disk material
more easily \citep[e.g.,][]{JOHNSTONE+98}, the massive Trapezium stars
likely truncate the disk size distribution.

One might expect to see some dependence of disk size or flux with
(projected) distance from $\theta^1$ Ori C, since disks in the cluster
outskirts would be less affected by photoionizing radiation.  
However, the crossing time within a
fraction of a parsec is substantially shorter than the age of the
region.  Indeed, the short disk lifetimes suggested by inferred
photoevaporation rates for free-free-dominated sources argue that
these objects have not been close to the Trapezium stars for longer
than $10^4$--$10^5$ years \cite[e.g.,][]{BALLY+98}.
Thus, cluster kinematics may
dilute any underlying effects caused by (current) proximity to the
Trapezium cluster.

Previous observations found that sources within
0.03 pc of $\theta^1$ Ori C seem to have a truncated sub-mm flux
distribution with respect to disks at larger cluster radii
\citep{MW09,MW10,MANN+14,EISNER+16}.   Figure \ref{fig:clusterrad}
shows sub-mm flux plotted against distance from $\theta^1$ Ori C for
our current study. The observations presented here confirm
the lack of high-submm-flux sources within 0.03 pc.  
These disks, with lower sub-mm fluxes, are within the EUV-dominated
regime of $\theta^1$ Ori C, where intense ionizing radiation likely
photoevaporated the outer disks of these objects
\citep[e.g.,][]{JOHNSTONE+98}.

We find a marginally significant
correlation between observed sub-mm flux
and cluster radius over the decade or so of projected distances from
$\theta^1$ Ori C probed with our sample. 
While EUV radiation likely dominates within 0.03 pc, FUV radiation
from $\theta^1$ Ori C is responsible for disk
photoevaporation at larger cluster radii.  Figure \ref{fig:clusterrad}
indicates that this FUV radiation leads to lower-mass disks at smaller
cluster radii.  This behavior is easily explained if external disk
photoevaporation from $\theta^1$ Ori C is stripping disk matter.  At
greater distances the FUV flux decreases and hence the
photoevaporation rate is lower.  It is also possible that tidal
stripping of disks by stellar encounters, which occurs more frequently
in the denser inner regions, contributes to the observed trend.

Disks that suffer more photoevaporation should also be smaller, since
matter at larger stellocentric radii is less tightly bound to the
system and hence more easily stripped.  Figure \ref{fig:sclusterrad}
shows that our observations are consistent with increasing disk size
with cluster radius, though we do not constrain the relationship with
high significance.  Note that even very close to $\theta^1$ Ori C,
the inner disks can survive (since the gravitational radius is $\ga 1$
AU), and hence near-IR excess would still be observed
\citep[see, e.g.,][]{RICHERT+15}.

\epsscale{1.0}
\begin{figure*}[tbhp]
\plottwo{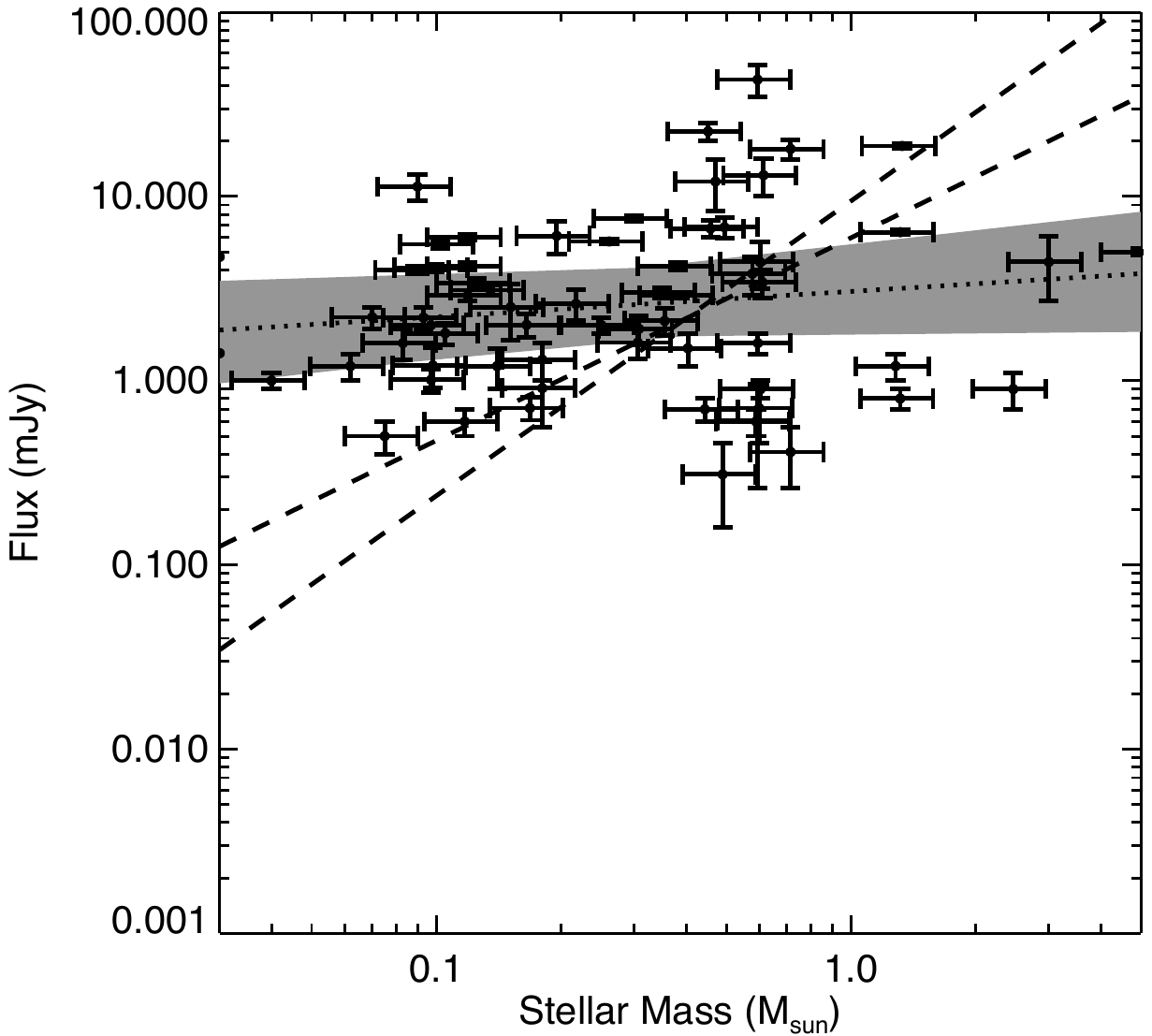}{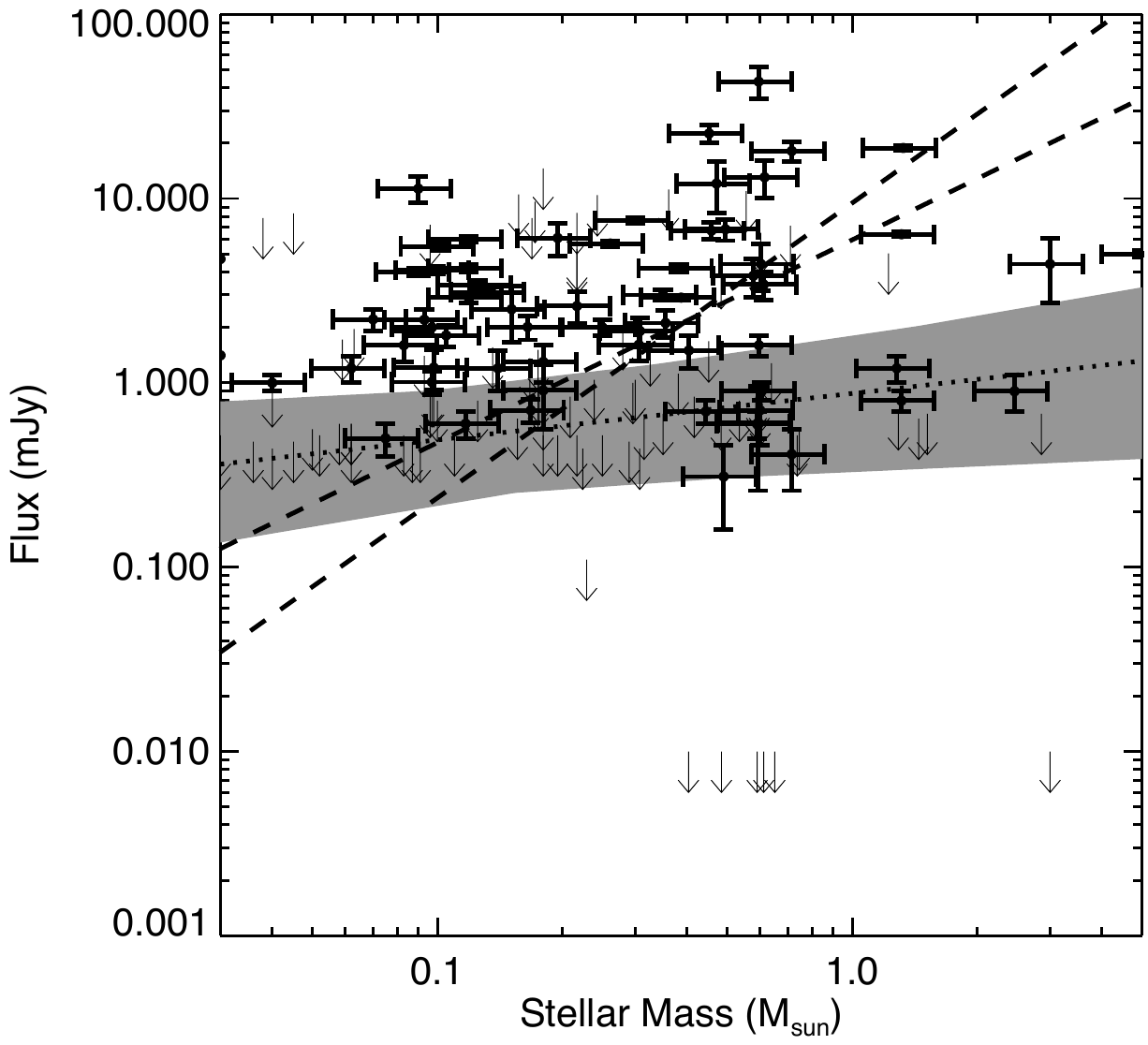}
\caption{Observed sub-mm flux from dusty disks plotted as a function
  of stellar mass.  Stellar masses are
  taken primarily from Fang et al. (in prep), who applied the
  evolutionary tracks of \citet{BARAFFE+15} to luminosities and
  effective temperatures determined from optical/near-IR spectra.
  Only objects for which stellar masses 
  have been determined are included, and we
  restrict our attention to (the vast majority of) stars less massive
  than 5 M$_{\odot}$.  We
  assume that stellar masses are uncertain by 20\%. In the
  right panel, we plot the same quantities, but include non-detected
  sources.  Regressions are shown for each
  panel, and the results are consistent.  When we include
  non-detections, the fitted trend is:
  log (Flux/mJy) $= (-0.06 \pm 0.30) + (0.25 \pm 0.15)$ log
  ($M_{\ast}/M_{\odot}$).
  For comparison we also show the relationships inferred for Taurus
  under different assumptions
  \citep[][]{ANDREWS+13,PASCUCCI+16,WARD-DUONG+17} as dashed lines in
  each panel.
\label{fig:fluxmass}}
\end{figure*}

\subsection{Disk Properties as a function of Stellar
  Mass \label{sec:stellar}}
Previous investigations of near-IR excess emission
showed the inner disk fraction for stars in Orion
to be largely independent of stellar age and mass, although there may
be a paucity of disks around very massive stars
\citep{HILLENBRAND+98,LADA+00}.  However this correlation has not been
well-studied in the ONC for massive disks as traced by mm/sub-mm emission.

Observations of Taurus and other low-density star-forming regions 
suggest a significant correlation between disk and stellar mass
\citep{ANDREWS+13,ANSDELL+16,PASCUCCI+16,BARENFELD+16}.
With our ONC sample we can explore whether or not sub-mm flux
correlates with stellar mass in the same way seen in these lower
density star-forming regions.

Stellar masses, determined spectroscopically, are available in the
literature for  about 60\% of our ALMA-detected sample
\citep{HILLENBRAND97,LUHMAN+00,SLESNICK+04,HILLENBRAND+13,INGRAHAM+14}. 
Our group has also been determining stellar masses for these and
additional cluster members using optical/near-IR spectroscopy (Fang et al. in
prep).  These new measurements are done in a uniform way, using
updated evolutionary tracks \citep[here we adopt those
of][]{BARAFFE+15}.  We use these instead of previous measurements where
possible, and supplement with stellar masses from the literature
when new measurements are not available.  We verify that the analysis
below gives consistent results with and without these supplemental
measurements.
We have stellar mass measurements for $\sim
70\%$ of our sub-mm detections (Table \ref{tab:detections}), and 65\%
of non-detected sources (Table \ref{tab:nondet}).   

We plot the sub-mm flux of detected sources (after removal of
free-free emission) versus stellar mass for this sub-sample in Figure
\ref{fig:fluxmass}.  In the right panel of Figure \ref{fig:fluxmass},
we make the same plot but include non-detected sources.
A weak correlation of sub-mm flux and stellar mass is seen. 

In contrast to our ONC results, in lower-density star-forming regions
a steeper trend is seen between disk flux and stellar mass
\citep[e.g.,][]{PASCUCCI+16}.  In Figure \ref{fig:fluxmass}, we plot the
stellar mass-disk flux relationship for Taurus, which, depending on
assumptions made, has a log-log slope varying from linear to 1.6
\citep{ANDREWS+13,PASCUCCI+16,WARD-DUONG+17}.
Even a linear slope is significantly 
steeper than allowed by the ONC data.  In Lupus, Cham I, and Upper
Sco, the relationship is even steeper than Taurus.  

\begin{figure}[tbp]
\plotone{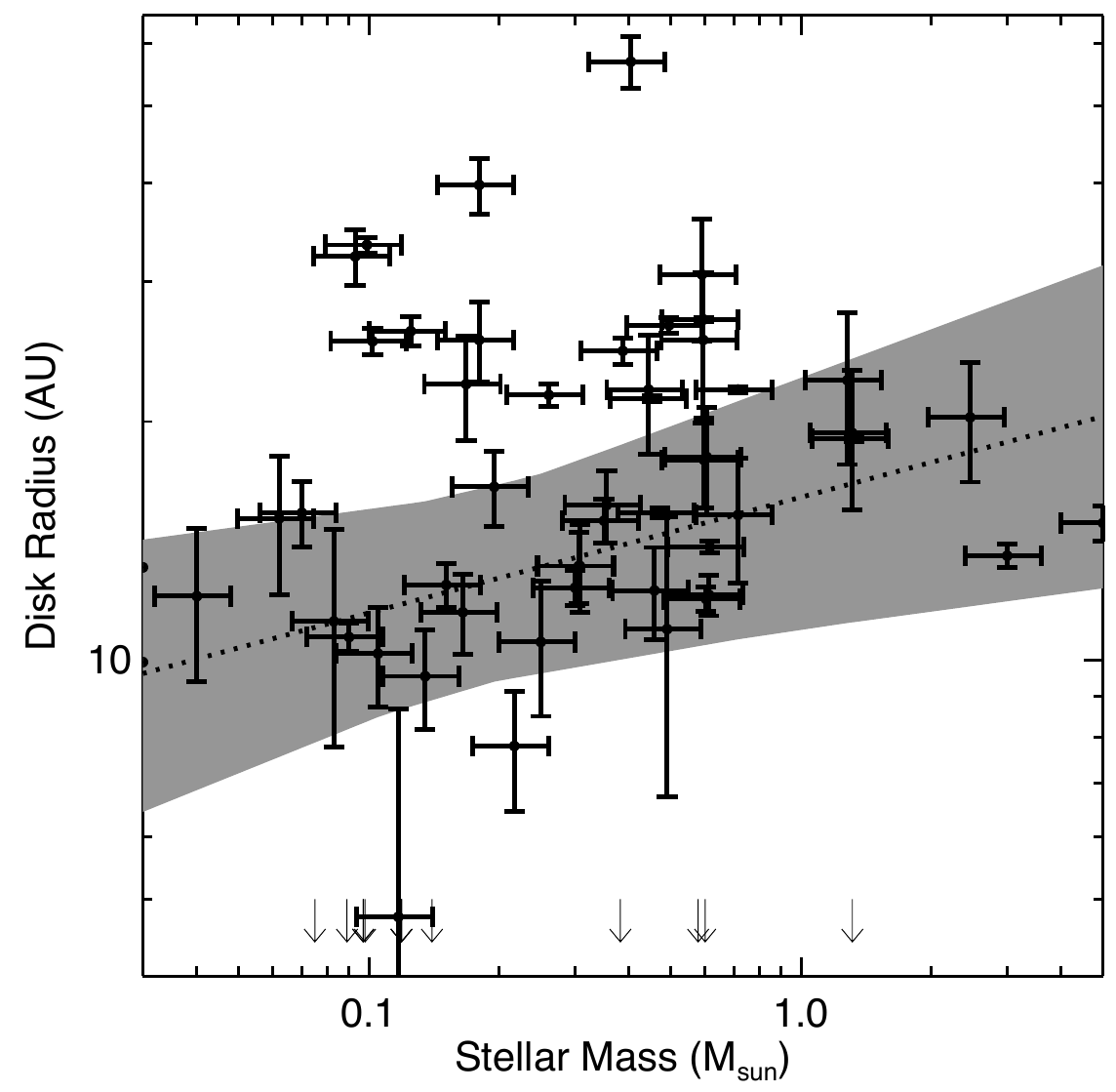}
\caption{Disk size (HWHM) for dusty disks as a function of stellar
  mass.  A linear regression to the data shows a positive correlation:
log (Radius/AU) = $(1.2 \pm 0.2) + (0.15 \pm 0.07)$ log
(M$_{\ast}$/M$_{\odot}$).
\label{fig:sizemass}}
\end{figure}

We also find a marginally significant ($\sim 2 \sigma$)
correlation of disk size and stellar mass (Figure
\ref{fig:sizemass}).  Disks around higher-mass stars tend to have
larger radii.  This may reflect an intrinsic relationship between disk
size and stellar mass, or it may result from the interplay between
external photoevporation and the gravitational binding energy of the
star.  Since the gravitational radius is $2G M_{\ast} / c_{\rm s}^2$,
photoevaporation will strip disks around lower-mass stars to smaller
stellocentric radii.

\begin{figure*}[tbp]
\plottwo{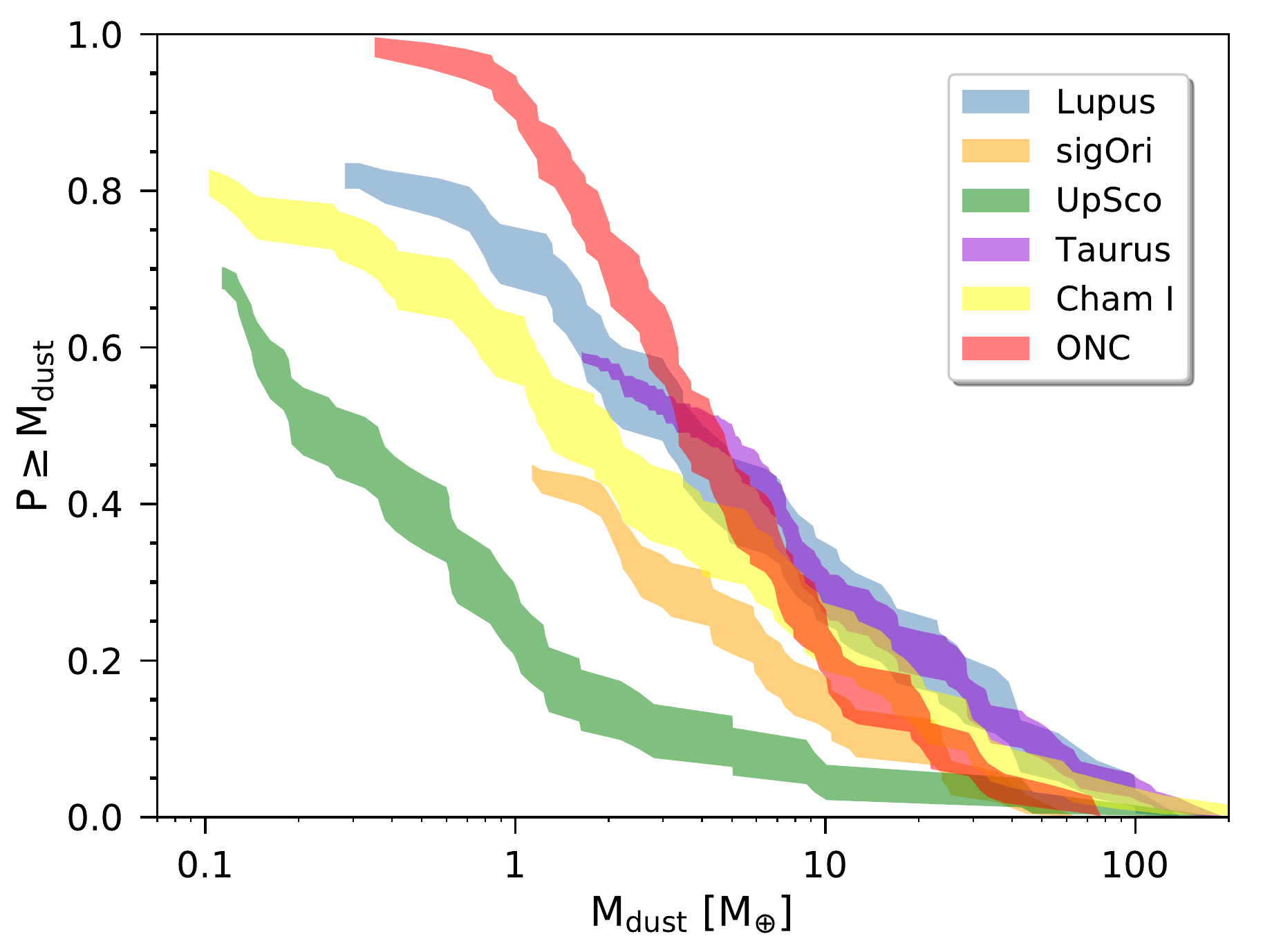}{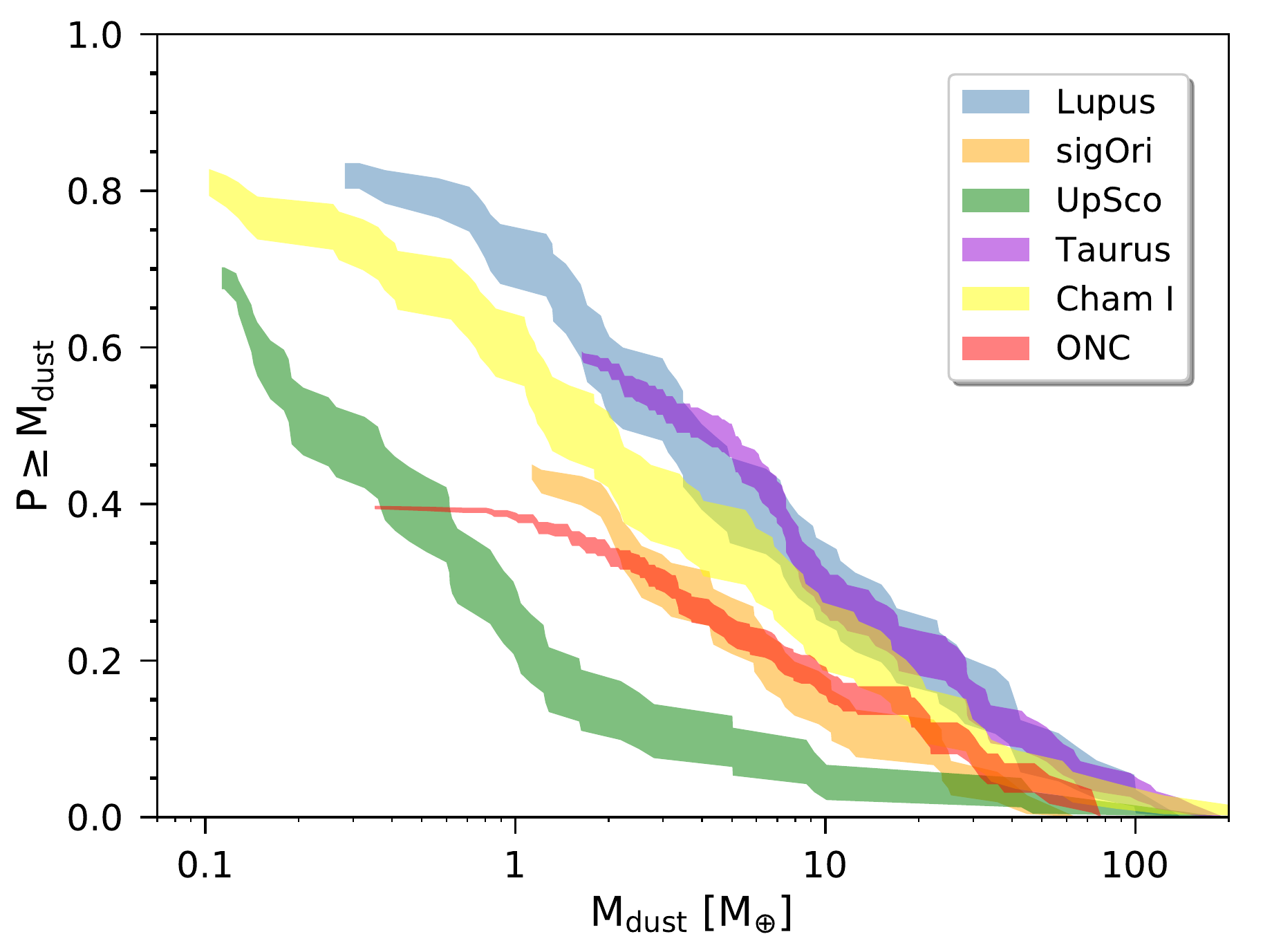}
\caption{Disk dust mass cumulative distribution for the ONC compared
  with previous work in Taurus \citep{ANDREWS+13}, Lupus
  \citep{ANSDELL+16}, Cham 1 \citep{PASCUCCI+16}, $\sigma$ Ori
  \citep{ANSDELL+17}, and Upper  Sco \citep{BARENFELD+16}.  
  The distributions and the 1$\sigma$ confidence intervals (shaded
  regions) were calculated using the Kaplan-Meier estimator.
  In the left panel, we use only
  detected, dusty disks in the ONC.  In the right panel we use both
  detections and censored non-detections for  the ONC. 
\label{fig:mdists}}
\end{figure*}

\subsection{Disk Mass Distribution \label{sec:mdist}}
Historically, disk masses have been inferred from mm and sub-mm
wavelength observations under the assumption of optically thin dust.
Since dust opacity tends to decrease with increasing wavelength
\citep[e.g.,][]{MN93}, disks are more likely to be optically thin in
the sub-mm and mm regime than at shorter wavelengths.  
At even longer, radio wavelengths, 
lower dust emissivity makes detection difficult, and free-free
emission from ionized gas can dominate.  Thus, the sub-mm is the
``sweet spot'' for measurements of dust continuum emission.

Assuming optically thin dust, continuum fluxes (less free-free
contributions) can be converted to disk dust masses:
\begin{equation}
M_{\rm dust} = \frac{S_{\rm \nu,dust} d^2} 
{\kappa_{\rm \nu,dust} B_{\nu}(T_{\rm dust})}.
\label{eq:dustmass}
\end{equation}
Here, $\nu$ is the observed frequency,
$S_{\rm \nu,dust}$ is the observed flux due to cool dust, $d$ is the distance 
to the source,
$\kappa_{\rm \nu, dust} = \kappa_0 (\nu / \nu_0)^{\beta}$ is the 
dust mass opacity,
$T_{\rm dust}$ is the dust temperature, and $B_{\nu}$ is the Planck
function\footnote{In the regime of interest for this calculation, the
  Planck function differs from the Rayleigh-Jeans approximation by
  $\sim 50\%$, and hence the full Planck function must be used.}. 
To use this formula, we make the following assumptions:
$d = 400$ pc; $\kappa_0=2$ cm$^{2}$ g$^{-1}$ at 1.3 
mm; $\beta=1.0$ \citep{HILDEBRAND83,BECKWITH+90}; and $T_{\rm dust} = 20$ K
\citep[e.g.,][]{AW05,CARPENTER02,WAW05}.  

As we discuss further below, the average disk temperature is somewhat
uncertain.  \citet{ANDREWS+13}
argue that $T_{\rm dust}$ should be a function of host star type, and they
find that 20 K is appropriate for M stars, which are the dominant
stellar constituent of the ONC \citep[e.g.,][]{HILLENBRAND97}.
Furthermore, \citet{TAZZARI+17} finds that 20 K is consistent
with detailed radiative transfer modeling of disks around low-mass
stars in Lupus.  However, these studies typically dealt with large
disks, and smaller disks will have higher average dust temperatures
\citep[e.g.,][]{HENDLER+17}.  External heating from massive stars 
may also lead to higher average disk temperatures
\citep[e.g.,][]{RICCI+17}.

Using this formula with $T_{\rm dust}=20$ K, 
we convert the measured 850 $\mu$m fluxes into
dust masses, and include these in Table \ref{tab:detections} and Figure
\ref{fig:fdist}.  The vast majority of detections have dust disk masses
$\la 1$ M$_{\Earth}$.   As a point of reference, giant planet cores
likely have masses $\ga 10$ M$_{\Earth}$ \citep[e.g.,][]{POLLACK+96}, 
and thus disks with only 1 M$_{\Earth}$ of solids would be unable to
form Jupiter-like planets.

We compare the simply-derived masses listed in Table
\ref{tab:detections} with those in other star-forming regions.  Mass
distributions, computed in the same way, 
are available for Taurus \citep{ANDREWS+13}, Lupus
\citep{ANSDELL+16}, Cham I \citep{PASCUCCI+16}, $\sigma$ Ori
\citep{ANSDELL+17}, and Upper Sco \citep{BARENFELD+16}.
As noted in Section \ref{sec:sizedist}, these regions span a range of
ages from $\sim 1$ to 10 Myr.  $\sigma$ Ori, which was not included
above because disk sizes are not available in the literature, has an
inferred age of 3--5 Myr \citep[][and references therein]{ANSDELL+17}.

Figure \ref{fig:mdists} shows the cumulative distribution functions of
disk masses in the ONC and lower-density star-forming regions.
Compared to Taurus, which has a similar age, the ONC lacks massive
disks and contains a larger number of low-mass disks.  The disk mass
distribution in the ONC also appears skewed to low masses compared
with Lupus and Cham I.  The ONC appears more similar to the older
$\sigma$ Ori region in terms of the disk mass distribution.

While one possible conclusion of the above analysis is that giant
planet formation is rare in the ONC, we believe it is more likely that
estimating mass using Equation \ref{eq:dustmass} is inaccurate.
The key assumption--optically thin dust--can easily break down in the
compact disks in our sample.   Furthermore, the assumption that the
average dust temperature is 20 K may break down since compact disks
are heated to higher average temperatures by their central stars.

To test the validity of Equation \ref{eq:dustmass}, we generated a
small set of radiative transfer models using RADMC-3D
\citep{DULLEMOND+12}. We adopted
a baseline model\footnote{Assumed disk parameters
are chosen to match those often-assumed in disk modeling studies
\citep[e.g.,][]{WOITKE+16}.}
 with inner radius of 0.1 AU, characteristic radius
($R_{\rm c}$) of 25 AU,
disk flaring of $H/R \propto R^{0.15}$, $H/R = 0.1$ at 100 AU, 
and a surface density profile, $\Sigma \propto R^{-1}$ up to $R_{\rm
  c}$, and falling off exponentially thereafter. 
For the central star, we assumed properties consistent with a
pre-main-sequence M star, $L_{\ast} = 0.5$ L$_{\odot}$ and $T_{\rm
  eff} = 3500$ K.   The dust grain sizes range from 0.05 $\mu$m to 10000
$\mu$m with power law slope of -3.5. 
This yields an opacity at 850 um of 3.5 cm$^2$ g$^{-1}$
 (consistent with the assumed opacity used in Equation \ref{eq:dustmass}).
Dust mass was varied between $10^{-2}$ M$_{\oplus}$ and $10^3$
M$_{\Earth}$.

\begin{figure}[tbp]
\plotone{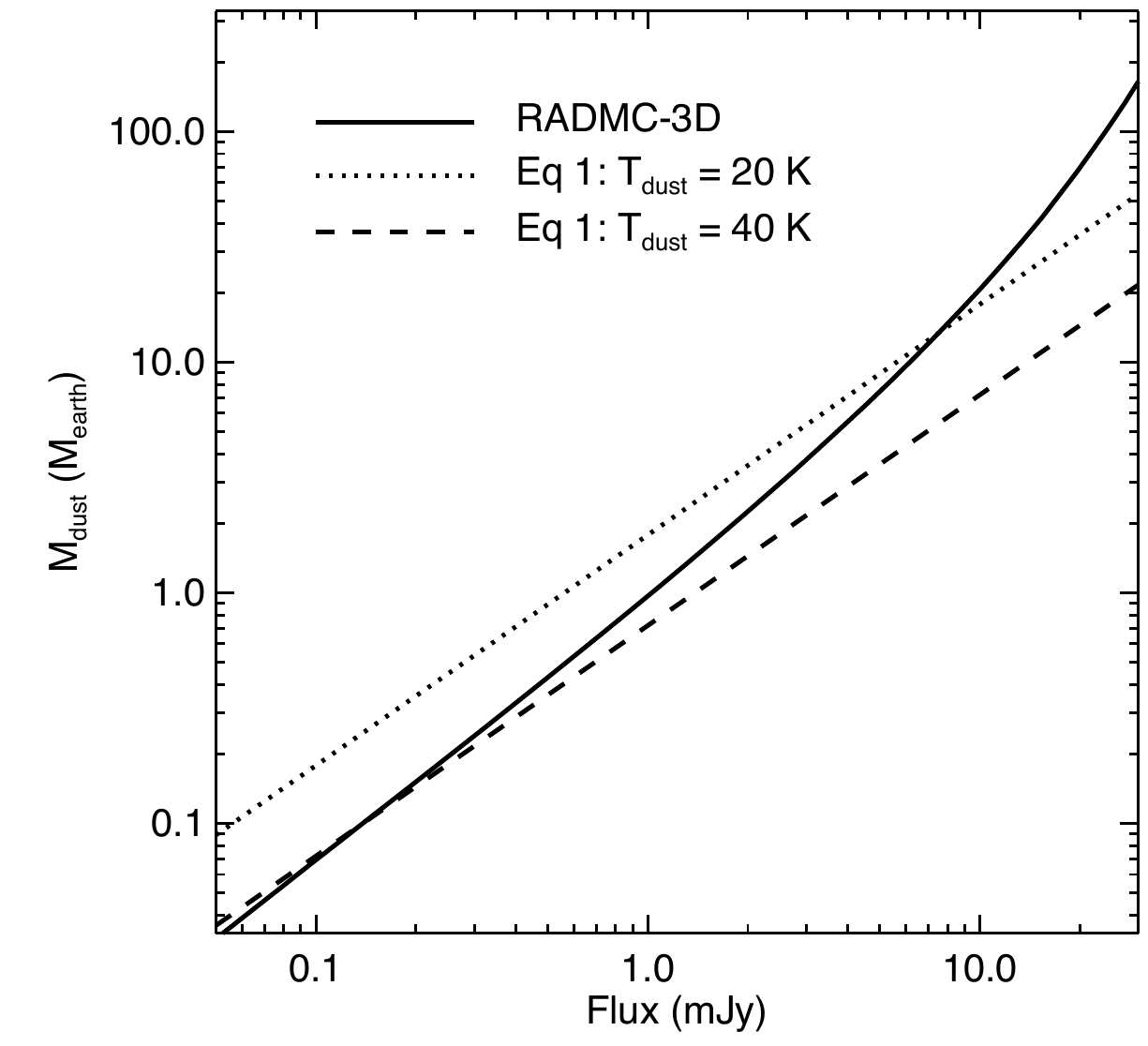}
\caption{Synthetic 850 $\mu$m fluxes produced by radiative transfer
  disk models with characteristic radii of 25 AU and a range of dust masses
  ({\it solid curve}).  Masses computed from Equation
  \ref{eq:dustmass} assuming $T_{\rm dust} = 20$ K and $T_{\rm dust} =
  40$ K are indicated as dotted and dashed lines, respectively.
\label{fig:comprt}}
\end{figure}

The 850 $\mu$m fluxes predicted by these models are shown in Figure
\ref{fig:comprt}.  For dust masses $\la 0.1$ M$_{\Earth}$,  disks
remain optically thin at most stellocentric radii.  However these
compact disks have average dust temperatures substantially higher than
the 20 K assumed above, using Equation \ref{eq:dustmass}.  As shown in
Figure \ref{fig:comprt}, 40 K is more appropriate.  Assuming $T_{\rm
  dust}=40$ K
and using Equation \ref{eq:dustmass} would {\it reduce} the dust
masses (for a given sub-mm flux value) by about a factor of 2 from
those listed in Table \ref{tab:detections}.

At higher disk masses, optical depth becomes significant, and hence
the masses computed using Equation \ref{eq:dustmass} are
underestimated.  For the brightest disks in our sample, with $F_{\nu}
> 10$ mJy, masses computed from Equation \ref{eq:dustmass} may be
underestimated by nearly an order or magnitude (Figure \ref{fig:comprt}).
As seen in Figure \ref{fig:comprt}, with an assumed $T_{\rm dust} = 20$ K, Equation
\ref{eq:dustmass} is within a factor of 2 of the radiative transfer
results for most sub-mm fluxes seen in our sample.

In future work we intend to compile broadband spectral energy
distributions for our sample (including new data to be obtained), and 
perform radiative transfer modeling for individual detections
to determine more accurate disk
masses.  These data will allow us to constrain many of
the parameters assumed in the simple models generated above.  These
additional constraints will improve the fidelity of computed disk dust
masses.

Even with such detailed modeling, one can only constrain the dust
mass, and not the mass in larger objects like planetesimals or giant
planet cores.  Collisional growth timescales for planetesimals are
significantly shorter than the $\sim 1$ Myr age of the ONC (and the
low-density star-forming regions discussed above), and hence
we expect a large fraction of the solid mass to be sequestered in
larger bodies in these systems. Comparison of inferred disk masses
with exoplanetary system masses leads to the same conclusion
\citep[e.g.,][]{NK14,PASCUCCI+16}.   Furthermore, evidence exists for the
formation of giant planet cores in our solar system
\citep{KRUIJER+17,DESCH+17}  and around other stars
\citep[e.g.,][]{ALMA+15,DONG+15,SE17} in less than 1 Myr.  

Thus, dust masses measured in ONC disks (and disks in other star
forming regions) may not relate directly to planet formation
potential.  In particular, the measurements described above may
underestimate the total solid inventory in disks and by extension,
the total disk masses.  The underestimate is, however, probably
smaller than an order of magnitude. If one assumes that the conversion
from sub-mm continuum flux to disk mass can be applied uniformly 
to the ensemble of disks in a given region (e.g., our ONC sample),
then scaling up eventually leads to gravitationally unstable (and
hence short-lived) disks at the high-mass end
\citep{EISNER+16,PASCUCCI+16}.

\section{Conclusions}
We presented ALMA observations of the center of the ONC, covering
approximately 225 cluster members.  We detected sub-mm emission from
104 sources, and spatially resolved 89 of these.
We removed sub-mm flux due to free-free emission in some objects,
leaving only flux attributable to dust emission.  12 objects detected
in our observations are consistent with the sub-mm flux originating
entirely from free-free emission.  Thus, we detect dusty disks around
92 sources.

Under standard assumptions of isothermal, optically thin
disks, the fluxes detected from dusty disks correspond to dust masses
ranging from 0.5 to 80
Earth masses.  We checked these estimates against more rigorous,
radiative transfer models
that include the effects of optical depth and varying dust temperature.
Comparison of the simple estimates with the more rigorous models suggest errors are
within about a factor of two for most objects, with the simple
estimates above ``true'' mass
values at low sub-mm fluxes and below true mass values at higher fluxes.

We measured the distribution of disk sizes, and found that disks in
this region are particularly compact.  Most disks have Gaussian HWHM
between 10 and 30 AU; no disks have radii $>60$ AU.  
Such compact disks are likely to be
significantly optically thick, and hotter than typically assumed.  The
compact disk sizes largely explain the discrepancy between simple disk
mass estimates and radiative transfer modeling.

Disk sizes are significantly smaller than those seen in
lower-density star-forming regions.   Disk masses in the ONC are also
significantly lower than in low-density star-forming regions of
similar age.  We suggest that the disks in the ONC have been affected
strongly by photoionzation, and possibly stellar encounters, in this
dense, high-mass star-forming region.

We argue that disks in the ONC have
been truncated by external photoevaporation, due to their proximity
to the massive star $\theta^1$ Ori C.  We found a 
correlation between disk flux and projected distance from 
$\theta^1$ Ori C, confirming that disk properties in this region are
influenced strongly by the environment.  The weak 
correlation of measured disk flux with stellar mass, compared to
steeper trends seen in lower-density star-forming regions of similar
age, likely reflects the impact of strong photoionization radiation
and high stellar density  in the ONC.

\vspace{0.2in}
This work was supported by NSF AAG grant 1311910.  J.E. is grateful to
Crystal Brogan for her extensive assistance with the data reduction.  I.P. also
acknowledges support from NSF AAG grant 151539. 
This paper makes use of the following ALMA data:
ADS/JAO.ALMA \#2015.1.00534.S. ALMA is a partnership of ESO
(representing its member states), NSF (USA) and NINS (Japan), together
with NRC (Canada), MOST and ASIAA (Taiwan), and KASI (Republic of
Korea), in cooperation with the Republic of Chile. The Joint ALMA
Observatory is operated by ESO, AUI/NRAO and NAOJ.
The National Radio Astronomy Observatory is a facility of the
National Science Foundation operated under cooperative agreement by
Associated Universities, Inc. The results reported herein benefitted
from collaborations and/or information exchange within NASA's Nexus
for Exoplanet System Science (NExSS) research coordination network
sponsored by NASA’s Science Mission Directorate.

\clearpage
\bibliographystyle{apj}
\bibliography{jae_ref}

\clearpage
\LongTables
\begin{deluxetable*}{lccccccc}
\tabletypesize{\scriptsize}
\tablewidth{0pt}
\tablecaption{Fluxes and Inferred Disk Masses for ALMA-detected Sources \label{tab:detections}}
\tablehead{\colhead{ID} & \colhead{$\alpha$} & \colhead{$\delta$} &
  \colhead{$M_{\ast}$} & \colhead{$F_{\rm \lambda 850 \mu m}$} & \colhead{$F_{\rm dust}$} &
  \colhead{$M_{\rm dust}$$^{\ast}$} & \colhead{$R_{\rm disk}$} \\
& (J2000) & (J2000) & (M$_{\odot}$) & (mJy) & (mJy) & ($M_{\Earth}$) & (AU)}
\startdata
HC273 &  5 35 13.41 &  -5 23 29.30 & -- &  5.6 $\pm$  0.4 &  5.6 $\pm$  0.4 &  9.8 $\pm$  0.7 & $15.1\pm 1.2$ \\
HC360 &  5 35 13.53 &  -5 23  4.50 &  0.10 &  2.0 $\pm$  0.2 &  2.0 $\pm$  0.2 &  3.5 $\pm$  0.3 & $< 5$ \\
HC192 &  5 35 13.59 &  -5 23 55.30 & -- & 12.2 $\pm$  2.5 & 12.5 $\pm$  4.8 & 21.8 $\pm$  8.4 & $16.8\pm 0.7$ \\
HC242 &  5 35 13.80 &  -5 23 40.20 & -- & 31.5 $\pm$  0.9 & 31.5 $\pm$  0.9 & 55.0 $\pm$  1.6 & $15.1\pm 0.5$ \\
139-320 &  5 35 13.92 &  -5 23 20.16 &  0.07 &  2.3 $\pm$  0.2 &  2.2 $\pm$  0.3 &  3.9 $\pm$  0.5 & $15.4\pm 1.4$ \\
142-301 &  5 35 14.15 &  -5 23  0.91 & -- &  6.8 $\pm$  0.3 &  2.5 $\pm$  2.0 &  4.5 $\pm$  3.5 & $37.7\pm 0.7$ \\
HC361 &  5 35 14.29 &  -5 23  4.30 & -- & 19.9 $\pm$  0.5 & 19.9 $\pm$  0.5 & 34.7 $\pm$  0.9 & $29.1\pm 0.3$ \\
HC345 &  5 35 14.32 &  -5 23  8.30 &  0.09 &  4.0 $\pm$  0.2 &  4.0 $\pm$  0.2 &  7.0 $\pm$  0.3 & $< 5$ \\
HC399 &  5 35 14.37 &  -5 22 54.10 & -- & 11.2 $\pm$  0.4 & 11.2 $\pm$  0.4 & 19.6 $\pm$  0.7 & $11.8\pm 0.6$ \\
HC391 &  5 35 14.39 &  -5 22 55.70 &  0.18 &  1.3 $\pm$  0.3 &  1.3 $\pm$  0.3 &  2.3 $\pm$  0.5 & $39.8\pm 3.2$ \\
HC189 &  5 35 14.53 &  -5 23 56.00 & -- & 46.3 $\pm$  1.9 & 46.3 $\pm$  1.9 & 80.8 $\pm$  3.3 & $43.3\pm 0.7$ \\
HC364 &  5 35 14.54 &  -5 23  3.70 &  0.12 &  4.2 $\pm$  0.2 &  4.2 $\pm$  0.2 &  7.3 $\pm$  0.3 & $< 5$ \\
HC276 &  5 35 14.66 &  -5 23 28.70 & -- &  0.7 $\pm$  0.1 &  0.7 $\pm$  0.1 &  1.2 $\pm$  0.2 & $18.4\pm 4.0$ \\
HC756/7 &  5 35 14.67 &  -5 22 38.60 & -- & 17.3 $\pm$  2.0 & 17.3 $\pm$  2.0 & 30.2 $\pm$  3.5 & $45.4\pm 2.1$ \\
HC411 &  5 35 14.70 &  -5 22 49.40 &  0.46 &  6.7 $\pm$  0.7 &  6.7 $\pm$  0.7 & 11.7 $\pm$  1.2 & $12.2\pm 1.6$ \\
147-323 &  5 35 14.72 &  -5 23 23.01 &  0.50 &  6.2 $\pm$  0.2 &  6.8 $\pm$  0.9 & 11.9 $\pm$  1.6 & $26.5\pm 0.6$ \\
148-305 &  5 35 14.80 &  -5 23  4.76 & -- &  0.7 $\pm$  0.2 &  0.5 $\pm$  0.3 &  0.9 $\pm$  0.5 & $15.8\pm 4.4$ \\
HC771 &  5 35 14.86 &  -5 22 44.10 & -- & 13.0 $\pm$  1.3 & 13.0 $\pm$  1.3 & 22.7 $\pm$  2.3 & $31.8\pm 2.2$ \\
HC714 &  5 35 14.88 &  -5 23  5.10 &  0.38 &  4.2 $\pm$  0.2 &  4.2 $\pm$  0.2 &  7.3 $\pm$  0.3 & $< 5$ \\
149-329 &  5 35 14.92 &  -5 23 29.05 &  0.17 &  0.8 $\pm$  0.1 &  0.7 $\pm$  0.1 &  1.2 $\pm$  0.2 & $22.3\pm 3.4$ \\
HC334 &  5 35 15.00 &  -5 23 14.30 & -- &  0.7 $\pm$  0.1 &  0.7 $\pm$  0.1 &  1.2 $\pm$  0.2 & $< 5$ \\
HC398 &  5 35 15.20 &  -5 22 54.40 &  1.28 &  1.2 $\pm$  0.2 &  1.2 $\pm$  0.2 &  2.1 $\pm$  0.3 & $22.5\pm 4.9$ \\
152-319 &  5 35 15.20 &  -5 23 18.81 &  0.10 &  2.2 $\pm$  0.1 &  1.8 $\pm$  0.2 &  3.2 $\pm$  0.4 & $10.2\pm 1.5$ \\
154-324 &  5 35 15.35 &  -5 23 24.11 & -- &  0.6 $\pm$  0.1 &  0.2 $\pm$  0.2 &  0.4 $\pm$  0.3 & $31.9\pm 4.7$ \\
154-240 &  5 35 15.38 &  -5 22 39.85 &  0.19 &  6.6 $\pm$  0.6 &  6.1 $\pm$  1.2 & 10.7 $\pm$  2.2 & $16.5\pm 1.8$ \\
HC223 &  5 35 15.44 &  -5 23 45.50 &  0.22 &  2.8 $\pm$  0.2 &  2.6 $\pm$  0.5 &  4.6 $\pm$  0.9 & $ 7.8\pm 1.3$ \\
HC413 &  5 35 15.49 &  -5 22 48.60 &  0.40 &  1.5 $\pm$  0.3 &  1.5 $\pm$  0.3 &  2.6 $\pm$  0.5 & $56.8\pm 4.3$ \\
155-338 &  5 35 15.51 &  -5 23 37.45 &  0.47 & 16.5 $\pm$  0.4 & 12.1 $\pm$  3.8 & 21.0 $\pm$  6.5 & $15.4\pm 0.2$ \\
HC389 &  5 35 15.64 &  -5 22 56.40 &  0.60 &  0.8 $\pm$  0.2 &  0.7 $\pm$  0.2 &  1.2 $\pm$  0.4 & $< 5$ \\
HC246 &  5 35 15.68 &  -5 23 39.10 &  1.33 & 18.8 $\pm$  0.5 & 18.8 $\pm$  0.5 & 32.8 $\pm$  0.9 & $19.0\pm 0.2$ \\
157-323$^{\dagger}$ &  5 35 15.72 &  -5 23 22.59 &  0.65 &  1.3 $\pm$  0.2 &  0.0 $\pm$  0.2 &  0.0 $\pm$  0.4 & $24.2\pm 2.9$ \\
158-327 &  5 35 15.79 &  -5 23 26.51 &  3.00 &  8.7 $\pm$  0.2 &  4.4 $\pm$  1.7 &  7.7 $\pm$  3.0 & $13.6\pm 0.5$ \\
HC336$^{\dagger}$ &  5 35 15.81 &  -5 23 14.30 & 18.91 &  6.6 $\pm$  0.2 &  0.0 $\pm$  0.3 &  0.0 $\pm$  0.5 & $14.6\pm 0.6$ \\
158-323$^{\dagger}$ &  5 35 15.83 &  -5 23 22.59 &  0.61 &  5.3 $\pm$  0.2 &  0.0 $\pm$  0.2 &  0.0 $\pm$  0.4 & $21.9\pm 0.9$ \\
HC291$^{\dagger}$ &  5 35 15.84 &  -5 23 25.60 &  0.17 &  4.4 $\pm$  0.2 &  1.5 $\pm$  1.8 &  2.6 $\pm$  3.1 & $14.1\pm 0.8$ \\
HC342 &  5 35 15.85 &  -5 23 11.00 &  5.00 &  5.0 $\pm$  0.2 &  5.0 $\pm$  0.2 &  8.7 $\pm$  0.3 & $14.9\pm 0.8$ \\
HC370 &  5 35 15.88 &  -5 23  2.00 & -- &  5.9 $\pm$  0.2 &  6.1 $\pm$  1.1 & 10.7 $\pm$  1.9 & $10.6\pm 0.6$ \\
HC447 &  5 35 15.89 &  -5 22 33.20 &  0.08 &  1.6 $\pm$  0.3 &  1.6 $\pm$  0.3 &  2.8 $\pm$  0.5 & $11.2\pm 3.4$ \\
159-338 &  5 35 15.90 &  -5 23 38.00 &  0.15 &  3.6 $\pm$  0.1 &  2.5 $\pm$  0.9 &  4.4 $\pm$  1.5 & $12.4\pm 0.8$ \\
159-350 &  5 35 15.96 &  -5 23 50.30 &  0.60 & 44.7 $\pm$  1.1 & 43.1 $\pm$  8.5 & 75.3 $\pm$ 14.8 & $26.9\pm 0.2$ \\
160-353 &  5 35 16.01 &  -5 23 53.00 &  0.58 &  4.8 $\pm$  0.2 &  3.8 $\pm$  0.9 &  6.7 $\pm$  1.6 & $< 5$ \\
161-324$^{\dagger}$ &  5 35 16.05 &  -5 23 24.35 & -- &  2.4 $\pm$  0.2 &  0.0 $\pm$  0.1 &  0.0 $\pm$  0.1 & $13.8\pm 1.4$ \\
HC350 &  5 35 16.06 &  -5 23  7.30 & -- &  4.6 $\pm$  0.2 &  2.0 $\pm$  0.9 &  3.5 $\pm$  1.7 & $27.0\pm 0.8$ \\
161-328 &  5 35 16.07 &  -5 23 27.81 & -- &  2.7 $\pm$  0.2 &  1.9 $\pm$  0.4 &  3.3 $\pm$  0.8 & $21.2\pm 1.1$ \\
HC401 &  5 35 16.08 &  -5 22 54.10 &  0.06 &  1.2 $\pm$  0.2 &  1.2 $\pm$  0.2 &  2.1 $\pm$  0.3 & $15.1\pm 3.0$ \\
161-314 &  5 35 16.10 &  -5 23 14.05 &  0.18 &  1.2 $\pm$  0.2 &  0.9 $\pm$  0.3 &  1.6 $\pm$  0.6 & $25.4\pm 2.9$ \\
HC354$^{\dagger}$ &  5 35 16.11 &  -5 23  6.80 & -- &  0.7 $\pm$  0.2 &  0.0 $\pm$  0.1 &  0.0 $\pm$  0.1 & $27.0\pm 0.8$ \\
HC393 &  5 35 16.14 &  -5 22 55.20 &  0.25 &  2.0 $\pm$  0.2 &  2.0 $\pm$  0.2 &  3.5 $\pm$  0.3 & $10.6\pm 2.0$ \\
163-317$^{\dagger}$ &  5 35 16.27 &  -5 23 16.51 &  0.40 &  3.5 $\pm$  0.2 &  0.0 $\pm$  0.1 &  0.0 $\pm$  0.1 & $16.7\pm 1.1$ \\
163-222 &  5 35 16.30 &  -5 22 21.50 &  0.09 &  2.2 $\pm$  0.3 &  2.2 $\pm$  0.3 &  3.8 $\pm$  0.5 & $32.3\pm 2.6$ \\
163-249 &  5 35 16.33 &  -5 22 49.01 &  0.31 &  2.1 $\pm$  0.2 &  1.9 $\pm$  0.3 &  3.3 $\pm$  0.6 & $13.1\pm 1.7$ \\
HC171 &  5 35 16.38 &  -5 24  3.40 &  0.59 &  0.9 $\pm$  0.2 &  0.6 $\pm$  0.3 &  1.1 $\pm$  0.6 & $25.4\pm 5.4$ \\
165-254 &  5 35 16.54 &  -5 22 53.70 & -- &  1.0 $\pm$  0.2 &  1.0 $\pm$  0.2 &  1.7 $\pm$  0.3 & $23.5\pm 3.5$ \\
HC165 &  5 35 16.58 &  -5 24  6.10 &  0.10 &  1.0 $\pm$  0.2 &  1.0 $\pm$  0.2 &  1.8 $\pm$  0.3 & $< 5$ \\
166-316$^{\dagger}$ &  5 35 16.61 &  -5 23 16.19 &  0.23 &  1.2 $\pm$  0.2 &  0.1 $\pm$  0.2 &  0.2 $\pm$  0.4 & $11.1\pm 2.5$ \\
HC280 &  5 35 16.66 &  -5 23 28.90 & -- &  0.9 $\pm$  0.1 &  0.9 $\pm$  0.1 &  1.6 $\pm$  0.2 & $13.8\pm 3.1$ \\
167-231 &  5 35 16.73 &  -5 22 31.30 &  0.12 &  3.4 $\pm$  0.2 &  3.4 $\pm$  0.2 &  5.9 $\pm$  0.3 & $26.0\pm 1.1$ \\
167-317$^{\dagger}$ &  5 35 16.74 &  -5 23 16.51 &  3.00 &  4.2 $\pm$  0.2 &  0.0 $\pm$  0.1 &  0.0 $\pm$  0.1 & $26.4\pm 0.8$ \\
168-328 &  5 35 16.77 &  -5 23 28.06 &  0.03 &  3.2 $\pm$  0.2 &  1.4 $\pm$  0.6 &  2.5 $\pm$  1.0 & $13.1\pm 0.9$ \\
168-326$^{\dagger}$ &  5 35 16.83 &  -5 23 25.91 & -- &  4.8 $\pm$  0.2 &  0.0 $\pm$  0.1 &  0.0 $\pm$  0.2 & $21.5\pm 0.9$ \\
HC397 &  5 35 16.91 &  -5 22 55.10 &  0.12 &  2.9 $\pm$  0.2 &  2.9 $\pm$  0.2 &  5.1 $\pm$  0.3 & $< 5$ \\
170-301 &  5 35 16.95 &  -5 23  0.91 &  0.61 &  3.4 $\pm$  0.2 &  3.4 $\pm$  0.6 &  6.0 $\pm$  1.0 & $12.1\pm 0.7$ \\
170-249 &  5 35 16.96 &  -5 22 48.51 &  0.09 & 10.3 $\pm$  0.3 & 11.3 $\pm$  1.9 & 19.7 $\pm$  3.2 & $10.7\pm 0.4$ \\
HC182 &  5 35 16.96 &  -5 23 59.50 & -- &  2.0 $\pm$  0.2 &  2.0 $\pm$  0.2 &  3.5 $\pm$  0.3 & $14.8\pm 1.9$ \\
170-337 &  5 35 16.97 &  -5 23 37.15 &  0.62 & 15.3 $\pm$  0.3 & 13.1 $\pm$  3.0 & 22.8 $\pm$  5.2 & $13.9\pm 0.2$ \\
171-340 &  5 35 17.04 &  -5 23 39.75 &  0.45 & 17.1 $\pm$  0.4 & 22.5 $\pm$  2.5 & 39.4 $\pm$  4.4 & $21.4\pm 0.2$ \\
HC259 &  5 35 17.07 &  -5 23 34.00 &  0.60 &  6.2 $\pm$  0.2 &  4.4 $\pm$  1.2 &  7.7 $\pm$  2.2 & $11.9\pm 0.4$ \\
173-341 &  5 35 17.32 &  -5 23 41.40 &  0.10 &  1.7 $\pm$  0.1 &  1.2 $\pm$  0.3 &  2.1 $\pm$  0.5 & $< 5$ \\
173-236 &  5 35 17.34 &  -5 22 35.81 &  0.71 & 14.5 $\pm$  0.3 & 18.1 $\pm$  2.2 & 31.6 $\pm$  3.8 & $22.0\pm 0.2$ \\
HC422 &  5 35 17.38 &  -5 22 45.80 &  0.12 &  6.0 $\pm$  0.2 &  6.0 $\pm$  0.2 & 10.5 $\pm$  0.3 & $< 5$ \\
HC180 &  5 35 17.39 &  -5 24  0.30 &  2.46 &  0.9 $\pm$  0.2 &  0.9 $\pm$  0.2 &  1.6 $\pm$  0.3 & $20.2\pm 3.5$ \\
HC313 &  5 35 17.47 &  -5 23 21.10 &  0.49 &  0.6 $\pm$  0.1 &  0.3 $\pm$  0.2 &  0.5 $\pm$  0.3 & $10.9\pm 4.2$ \\
175-251 &  5 35 17.47 &  -5 22 51.26 &  0.35 &  2.2 $\pm$  0.2 &  2.1 $\pm$  0.3 &  3.7 $\pm$  0.6 & $15.7\pm 1.7$ \\
175-355 &  5 35 17.54 &  -5 23 55.05 & -- &  1.2 $\pm$  0.1 &  1.2 $\pm$  0.1 &  2.1 $\pm$  0.2 & $20.5\pm 2.4$ \\
176-325$^{\dagger}$ &  5 35 17.55 &  -5 23 24.96 &  0.48 &  2.3 $\pm$  0.1 &  0.0 $\pm$  0.1 &  0.0 $\pm$  0.1 & $31.9\pm 1.2$ \\
HC388 &  5 35 17.56 &  -5 22 56.80 &  0.60 &  1.6 $\pm$  0.2 &  1.6 $\pm$  0.2 &  2.8 $\pm$  0.3 & $17.9\pm 2.3$ \\
177-341W &  5 35 17.66 &  -5 23 41.00 & -- & 10.9 $\pm$  0.3 &  2.8 $\pm$  2.7 &  4.9 $\pm$  4.6 & $29.5\pm 0.3$ \\
HC234 &  5 35 17.77 &  -5 23 42.60 &  0.14 &  1.2 $\pm$  0.3 &  1.2 $\pm$  0.3 &  2.1 $\pm$  0.5 & $< 5$ \\
HC332 &  5 35 17.82 &  -5 23 15.60 &  0.39 &  2.9 $\pm$  0.1 &  2.9 $\pm$  0.1 &  5.1 $\pm$  0.2 & $24.5\pm 0.9$ \\
178-258 &  5 35 17.84 &  -5 22 58.15 &  0.26 &  5.7 $\pm$  0.2 &  5.7 $\pm$  0.2 & 10.0 $\pm$  0.3 & $21.6\pm 0.7$ \\
179-354 &  5 35 17.96 &  -5 23 53.50 &  0.04 &  1.0 $\pm$  0.1 &  1.0 $\pm$  0.1 &  1.7 $\pm$  0.2 & $12.1\pm 2.6$ \\
180-331 &  5 35 18.03 &  -5 23 30.80 & -- &  4.4 $\pm$  0.2 &  1.5 $\pm$  1.0 &  2.6 $\pm$  1.7 & $15.0\pm 0.6$ \\
HC174 &  5 35 18.04 &  -5 24  3.10 &  0.30 &  7.6 $\pm$  0.3 &  7.6 $\pm$  0.3 & 13.3 $\pm$  0.5 & $12.4\pm 0.6$ \\
181-247 &  5 35 18.08 &  -5 22 47.10 & -- &  4.1 $\pm$  0.2 &  4.7 $\pm$  0.7 &  8.2 $\pm$  1.2 & $15.4\pm 0.8$ \\
HC177 &  5 35 18.08 &  -5 24  1.20 &  0.14 &  3.1 $\pm$  0.2 &  3.1 $\pm$  0.2 &  5.4 $\pm$  0.3 & $ 9.6\pm 1.4$ \\
182-316 &  5 35 18.19 &  -5 23 31.55 &  0.35 &  3.0 $\pm$  0.2 &  3.0 $\pm$  0.2 &  5.2 $\pm$  0.3 & $15.0\pm 1.0$ \\
HC253 &  5 35 18.21 &  -5 23 35.90 &  1.32 &  6.4 $\pm$  0.2 &  6.4 $\pm$  0.2 & 11.2 $\pm$  0.3 & $< 5$ \\
HC331 &  5 35 18.25 &  -5 23 15.70 &  0.31 &  2.0 $\pm$  0.1 &  1.6 $\pm$  0.3 &  2.8 $\pm$  0.5 & $13.1\pm 1.4$ \\
HC436$^{\dagger}$ &  5 35 18.38 &  -5 22 37.50 &  0.59 &  1.8 $\pm$  0.1 &  0.0 $\pm$  0.1 &  0.0 $\pm$  0.1 & $ 6.7\pm 1.4$ \\
HC278 &  5 35 18.50 &  -5 23 29.30 &  0.08 &  0.5 $\pm$  0.1 &  0.5 $\pm$  0.1 &  0.9 $\pm$  0.2 & $< 5$ \\
187-314 &  5 35 18.68 &  -5 23 14.01 &  0.61 &  0.9 $\pm$  0.1 &  0.9 $\pm$  0.1 &  1.6 $\pm$  0.2 & $18.0\pm 2.8$ \\
HC191 &  5 35 18.68 &  -5 23 56.50 &  0.12 &  0.6 $\pm$  0.1 &  0.6 $\pm$  0.1 &  1.0 $\pm$  0.2 & $ 4.8\pm 3.9$ \\
HC713 &  5 35 18.71 &  -5 22 56.90 &  1.31 &  0.8 $\pm$  0.1 &  0.8 $\pm$  0.1 &  1.4 $\pm$  0.2 & $19.4\pm 3.9$ \\
HC482 &  5 35 18.85 &  -5 22 23.10 &  0.10 &  5.5 $\pm$  0.3 &  5.5 $\pm$  0.3 &  9.6 $\pm$  0.5 & $25.2\pm 0.9$ \\
189-329 &  5 35 18.87 &  -5 23 28.85 &  0.17 &  2.1 $\pm$  0.1 &  2.0 $\pm$  0.3 &  3.5 $\pm$  0.5 & $11.5\pm 1.3$ \\
HC352 &  5 35 18.88 &  -5 23  7.20 & -- &  1.7 $\pm$  0.1 &  1.7 $\pm$  0.1 &  3.0 $\pm$  0.2 & $< 5$ \\
190-251 &  5 35 19.03 &  -5 22 50.65 &  0.59 &  0.6 $\pm$  0.1 &  0.6 $\pm$  0.1 &  1.0 $\pm$  0.2 & $30.6\pm 5.4$ \\
191-350 &  5 35 19.06 &  -5 23 49.50 &  0.71 &  0.8 $\pm$  0.1 &  0.4 $\pm$  0.2 &  0.7 $\pm$  0.3 & $15.2\pm 2.7$ \\
HC351 &  5 35 19.07 &  -5 23  7.50 &  0.10 &  4.1 $\pm$  0.2 &  4.1 $\pm$  0.2 &  7.2 $\pm$  0.3 & $33.4\pm 0.8$ \\
HC288 &  5 35 19.12 &  -5 23 27.10 &  0.44 &  0.7 $\pm$  0.1 &  0.7 $\pm$  0.1 &  1.2 $\pm$  0.2 & $22.0\pm 3.7$ \\
191-232 &  5 35 19.13 &  -5 22 31.20 & -- &  1.1 $\pm$  0.1 &  1.1 $\pm$  0.1 &  1.9 $\pm$  0.2 & $34.5\pm 2.7$ \\
HC366 &  5 35 19.63 &  -5 23  3.60 &  0.03 &  4.7 $\pm$  0.2 &  4.7 $\pm$  0.2 &  8.2 $\pm$  0.3 & $ 9.9\pm 0.9$ \\
198-222 &  5 35 19.82 &  -5 22 21.55 & -- &  2.6 $\pm$  0.3 &  2.7 $\pm$  0.5 &  4.7 $\pm$  0.9 & $15.3\pm 2.8$ \\
\enddata
\tablecomments{Stellar masses, where available, are taken from a new
  study by Fang et al. (in prep), and from the
  literature 
\citep{HILLENBRAND97,LUHMAN+00,SLESNICK+04,HILLENBRAND+13,INGRAHAM+14}.
Disk sizes are HWHM major axes of Gaussian fits, after deconvolution of the
synthesized beam.  All uncertainties listed in the table are 1$\sigma$ errors.
$^{\ast}$--Dust masses are computed from the measured dust fluxes using Equation
\ref{eq:dustmass} and assuming $T_{\rm dust} = 20$ K.  These masses
are not accurate (see Figure \ref{fig:comprt} and discussion in
Section \ref{sec:mdist}), and are listed only
to give a rough idea of how flux and mass are related.
$^{\dagger}$--After removal of free-free emission, the dust masses for
these objects are consistent with zero.  While they were detected with
ALMA, the detection traces free-free emission, and no significant
emission from dust is detected.}
\end{deluxetable*}

\clearpage
\begin{deluxetable*}{lccc}
\tabletypesize{\scriptsize}
\tablewidth{0pt}
\tablecaption{Upper Limits for Non-Detected Sources \label{tab:nondet}}
\tablehead{\colhead{ID} & \colhead{$\alpha$} & \colhead{$\delta$} &
\colhead{$F_{\rm \lambda 850 \mu m}$} \\
& (J2000) & (J2000) & (mJy)}
\startdata
HC190 &  5 35 13.23 &  -5 23 55.50 & $<$10.0 \\
HC301 &  5 35 13.26 &  -5 23 22.80 & $<$ 1.3 \\
HC380 &  5 35 13.29 &  -5 22 57.90 & $<$ 1.7 \\
133-353 &  5 35 13.30 &  -5 23 53.05 & $<$ 8.3 \\
HC471 &  5 35 13.37 &  -5 22 26.20 & $<$ 9.6 \\
HC198 &  5 35 13.38 &  -5 23 53.20 & $<$ 7.8 \\
HC240 &  5 35 13.45 &  -5 23 40.40 & $<$ 4.2 \\
HC266 &  5 35 13.53 &  -5 23 30.90 & $<$ 1.4 \\
HC178 &  5 35 13.55 &  -5 23 59.70 & $<$14.6 \\
HC222 &  5 35 13.68 &  -5 23 45.40 & $<$11.6 \\
HC483 &  5 35 13.75 &  -5 22 22.00 & $<$11.2 \\
HC254 &  5 35 13.86 &  -5 23 35.00 & $<$ 1.5 \\
HC451 &  5 35 13.97 &  -5 22 31.90 & $<$ 8.4 \\
HC247 &  5 35 14.05 &  -5 23 38.50 & $<$ 2.0 \\
HC438 &  5 35 14.09 &  -5 22 36.60 & $<$ 7.8 \\
HC458 &  5 35 14.31 &  -5 22 30.70 & $<$11.0 \\
HC448 &  5 35 14.36 &  -5 22 32.80 & $<$21.5 \\
HC439 &  5 35 14.37 &  -5 22 36.10 & $<$ 8.8 \\
HC300 &  5 35 14.40 &  -5 23 23.10 & $<$ 0.5 \\
HC258 &  5 35 14.40 &  -5 23 33.70 & $<$ 0.8 \\
HC759 &  5 35 14.50 &  -5 22 29.40 & $<$28.2 \\
HC784 &  5 35 14.50 &  -5 22 38.78 & $<$ 8.2 \\
HC193 &  5 35 14.53 &  -5 23 55.10 & $<$ 7.5 \\
HC209 &  5 35 14.57 &  -5 23 50.80 & $<$ 6.5 \\
HC443 &  5 35 14.66 &  -5 22 33.80 & $<$11.0 \\
HC369 &  5 35 14.67 &  -5 23  1.90 & $<$ 0.7 \\
HC755 &  5 35 14.71 &  -5 22 35.50 & $<$ 8.8 \\
HC464 &  5 35 14.73 &  -5 22 29.80 & $<$15.9 \\
HC220 &  5 35 14.82 &  -5 23 46.50 & $<$ 1.6 \\
HC773 &  5 35 14.82 &  -5 22 23.20 & $<$10.5 \\
HC324 &  5 35 14.84 &  -5 23 16.00 & $<$ 0.5 \\
HC453 &  5 35 14.87 &  -5 22 31.70 & $<$10.4 \\
HC431 &  5 35 14.92 &  -5 22 39.10 & $<$ 6.5 \\
HC245 &  5 35 14.95 &  -5 23 39.30 & $<$ 0.8 \\
150-231 &  5 35 15.02 &  -5 22 31.11 & $<$ 7.2 \\
HC195 &  5 35 15.04 &  -5 23 54.50 & $<$ 2.0 \\
HC373 &  5 35 15.04 &  -5 23  1.10 & $<$ 0.6 \\
HC298 &  5 35 15.07 &  -5 23 23.40 & $<$ 0.5 \\
HC219 &  5 35 15.16 &  -5 23 46.70 & $<$ 1.1 \\
HC359 &  5 35 15.18 &  -5 23  5.00 & $<$ 0.6 \\
HC437 &  5 35 15.21 &  -5 22 36.70 & $<$ 5.0 \\
HC478 &  5 35 15.21 &  -5 22 24.10 & $<$ 7.1 \\
HC211 &  5 35 15.25 &  -5 23 49.80 & $<$ 1.1 \\
HC386 &  5 35 15.27 &  -5 22 56.70 & $<$ 0.8 \\
HC299 &  5 35 15.30 &  -5 23 23.20 & $<$ 0.5 \\
HC310 &  5 35 15.35 &  -5 23 21.40 & $<$ 0.5 \\
HC476 &  5 35 15.35 &  -5 22 25.20 & $<$ 5.3 \\
154-225 &  5 35 15.37 &  -5 22 25.35 & $<$ 5.0 \\
HC261 &  5 35 15.38 &  -5 23 33.40 & $<$ 0.5 \\
HC327 &  5 35 15.54 &  -5 23 15.80 & $<$ 0.6 \\
HC419 &  5 35 15.55 &  -5 22 46.40 & $<$ 1.2 \\
HC274 &  5 35 15.56 &  -5 23 29.60 & $<$ 0.5 \\
HC378 &  5 35 15.60 &  -5 22 58.90 & $<$ 0.6 \\
HC172 &  5 35 15.62 &  -5 24  3.10 & $<$ 1.3 \\
HC236 &  5 35 15.70 &  -5 23 41.90 & $<$ 0.5 \\
HC248 &  5 35 15.76 &  -5 23 38.40 & $<$ 1.7 \\
HC344 &  5 35 15.77 &  -5 23  9.90 & $<$ 0.7 \\
158-326 &  5 35 15.81 &  -5 23 25.51 & $<$ 0.9 \\
HC340 &  5 35 15.81 &  -5 23 12.00 & $<$ 0.8 \\
HC420 &  5 35 15.84 &  -5 22 45.90 & $<$ 0.8 \\
HC769 &  5 35 15.96 &  -5 22 41.10 & $<$ 0.8 \\
HC304 &  5 35 15.97 &  -5 23 22.70 & $<$ 0.8 \\
HC303 &  5 35 16.10 &  -5 23 23.20 & $<$ 0.6 \\
HC768 &  5 35 16.14 &  -5 22 45.10 & $<$ 0.8 \\
HC435 &  5 35 16.20 &  -5 22 37.50 & $<$ 0.8 \\
HC758 &  5 35 16.24 &  -5 22 24.30 & $<$ 1.0 \\
HC317 &  5 35 16.24 &  -5 23 19.10 & $<$ 0.6 \\
HC479 &  5 35 16.31 &  -5 22 24.00 & $<$ 1.0 \\
HC292 &  5 35 16.35 &  -5 23 25.30 & $<$ 0.6 \\
HC485 &  5 35 16.38 &  -5 22 22.30 & $<$ 1.0 \\
HC341 &  5 35 16.41 &  -5 23 11.50 & $<$ 0.5 \\
HC309 &  5 35 16.46 &  -5 23 23.00 & $<$ 0.6 \\
165-235 &  5 35 16.48 &  -5 22 35.16 & $<$ 0.8 \\
HC390 &  5 35 16.49 &  -5 22 56.50 & $<$ 0.6 \\
166-406 &  5 35 16.57 &  -5 24  6.00 & $<$ 1.0 \\
166-250 &  5 35 16.59 &  -5 22 50.36 & $<$ 0.7 \\
HC293 &  5 35 16.73 &  -5 23 25.20 & $<$ 0.7 \\
HC170 &  5 35 16.77 &  -5 24  4.30 & $<$ 0.9 \\
168-235 &  5 35 16.81 &  -5 22 34.71 & $<$ 0.8 \\
HC235 &  5 35 16.84 &  -5 23 42.30 & $<$ 0.4 \\
HC349 &  5 35 16.87 &  -5 23  7.10 & $<$ 0.5 \\
169-338 &  5 35 16.88 &  -5 23 38.10 & $<$ 1.2 \\
HC484 &  5 35 16.90 &  -5 22 22.50 & $<$ 0.9 \\
HC441 &  5 35 16.91 &  -5 22 35.20 & $<$ 0.7 \\
HC450 &  5 35 17.01 &  -5 22 33.10 & $<$ 0.7 \\
HC410 &  5 35 17.12 &  -5 22 50.10 & $<$ 1.0 \\
HC315 &  5 35 17.16 &  -5 23 20.40 & $<$ 0.5 \\
HC330 &  5 35 17.24 &  -5 23 16.60 & $<$ 0.5 \\
HC493 &  5 35 17.34 &  -5 22 21.20 & $<$ 0.9 \\
174-305 &  5 35 17.37 &  -5 23  4.86 & $<$ 0.5 \\
HC237 &  5 35 17.41 &  -5 23 41.80 & $<$ 0.5 \\
HC469 &  5 35 17.58 &  -5 22 27.80 & $<$ 0.6 \\
176-252 &  5 35 17.64 &  -5 22 51.66 & $<$ 0.7 \\
177-341E &  5 35 17.73 &  -5 23 41.10 & $<$ 1.0 \\
HC333 &  5 35 17.74 &  -5 23 14.90 & $<$ 0.5 \\
HC462 &  5 35 17.76 &  -5 22 31.00 & $<$ 0.5 \\
HC230 &  5 35 17.79 &  -5 23 44.20 & $<$ 0.5 \\
HC367 &  5 35 17.87 &  -5 23  3.10 & $<$ 0.5 \\
HC425 &  5 35 17.95 &  -5 22 45.50 & $<$ 0.6 \\
HC256 &  5 35 17.97 &  -5 23 35.50 & $<$ 0.4 \\
HC372 &  5 35 18.08 &  -5 23  1.80 & $<$ 0.5 \\
HC221 &  5 35 18.21 &  -5 23 46.30 & $<$ 0.4 \\
HC348 &  5 35 18.28 &  -5 23  7.50 & $<$ 0.5 \\
183-405 &  5 35 18.33 &  -5 24  4.85 & $<$ 1.0 \\
HC430 &  5 35 18.40 &  -5 22 40.00 & $<$ 0.4 \\
HC185 &  5 35 18.51 &  -5 23 57.80 & $<$ 0.6 \\
HC765 &  5 35 18.53 &  -5 23 47.90 & $<$ 0.4 \\
HC384 &  5 35 18.53 &  -5 22 58.10 & $<$ 0.6 \\
HC463 &  5 35 18.58 &  -5 22 31.00 & $<$ 0.5 \\
HC311 &  5 35 18.97 &  -5 23 22.00 & $<$ 0.4 \\
HC357 &  5 35 19.11 &  -5 23  6.30 & $<$ 0.7 \\
HC444 &  5 35 19.14 &  -5 22 34.60 & $<$ 0.5 \\
HC408 &  5 35 19.22 &  -5 22 50.70 & $<$ 0.6 \\
HC356 &  5 35 19.38 &  -5 23  6.50 & $<$ 0.6 \\
HC491 &  5 35 19.47 &  -5 22 21.80 & $<$ 1.0 \\
HC728 &  5 35 19.51 &  -5 23 39.70 & $<$ 0.6 \\
HC188 &  5 35 19.62 &  -5 23 57.30 & $<$ 0.8 \\
HC446 &  5 35 19.68 &  -5 22 34.20 & $<$ 0.7 \\
HC210 &  5 35 19.86 &  -5 23 51.60 & $<$ 1.0 \\
HC176 &  5 35 19.93 &  -5 24  2.60 & $<$ 1.4 \\
HC452 &  5 35 19.98 &  -5 22 32.80 & $<$ 1.4 \\
HC766 &  5 35 20.00 &  -5 23 28.80 & $<$ 1.4 \\
HC474 &  5 35 20.03 &  -5 22 26.50 & $<$ 1.6 \\
\enddata
\tablecomments{These are 4$\sigma$ limits, where the rms is calculated
  locally toward each source position.}
\end{deluxetable*}

\end{document}